\documentclass[fleqn,usenatbib]{mnras}

\usepackage{newtxtext,newtxmath}

\usepackage[T1]{fontenc}

\DeclareRobustCommand{\VAN}[3]{#2}
\let\VANthebibliography\thebibliography
\def\thebibliography{\DeclareRobustCommand{\VAN}[3]{##3}\VANthebibliography}

\usepackage{graphicx}	
\usepackage{amsmath}	
\usepackage{lipsum}  

\title[Theoretical uncertainties in cluster counts]{The FLAMINGO Project: An assessment of the systematic errors in the predictions of models for galaxy cluster counts used to infer cosmological parameters}

\author[Kugel et al.]{
Roi Kugel,$^{1}$\thanks{E-mail: kugel@strw.leidenuniv.nl}
Joop Schaye,$^{1}$
Matthieu Schaller,$^{2,1}$
Victor J. Forouhar Moreno,$^{1}$
Robert J. McGibbon$^{1}$
\\
$^{1}$Leiden Observatory, Leiden University, PO Box 9513, 2300 RA Leiden, the Netherlands\\
$^{2}$Lorentz Institute for Theoretical Physics, Leiden University, PO box 9506, 2300 RA Leiden, the Netherlands\\
}

\date{Accepted XXX. Received YYY; in original form ZZZ}

\pubyear{2015}

\begin{document}
\label{firstpage}
\pagerange{\pageref{firstpage}--\pageref{lastpage}}
\maketitle

\begin{abstract} 
Galaxy cluster counts have historically been important for the measurement of cosmological parameters and upcoming surveys will greatly reduce the statistical errors. To exploit the potential of current and future cluster surveys, theoretical uncertainties on the predicted abundance must be smaller than the statistical errors. Models used to predict cluster counts typically combine a model for the dark matter only (DMO) halo mass function (HMF) with an observable – mass relation that is assumed to be a power-law with lognormal scatter. We use the FLAMINGO suite of cosmological hydrodynamical simulations to quantify the biases in the cluster counts and cosmological parameters resulting from the different ingredients of conventional models. For the observable mass proxy we focus on the Compton-Y parameter quantifying the thermal Sunyaev-Zel'dovich effect, which is expected to result in cluster samples that are relatively close to mass-selected samples. We construct three mock samples based on existing (Planck and SPT) and upcoming (Simons Observatory) surveys. We ignore measurement uncertainties and compare the biases in the counts and inferred cosmological parameters to each survey's Poisson errors. We find that widely used models for the DMO HMF differ significantly from each other and from the DMO version of FLAMINGO, leading to significant biases for all three surveys. For upcoming surveys, dramatic improvements are needed for all additional model ingredients, i.e. the functional forms of the fits to the observable-mass scaling relation and the associated scatter, the priors on the scaling relation and the prior on baryonic effects associated with feedback processes on the HMF.
\end{abstract}




\section{Introduction}
The standard model of cosmology, $\Lambda$CDM, consisting of a universe filled mostly by dark energy and cold dark matter, has so far been successful at describing observations, but cracks may have started to appear. When comparing the values of cosmological parameters inferred from the cosmic microwave background \citep{Planck2020} with the same values obtained from local measurements \citep[e.g.][]{Ries2022,KIDS2021,DESYR3} tensions have started to pop up. These tensions exist in particular for the Hubble constant, $H_0$, and for the clustering parameter $\sigma_8$. With the advent of surveys like Euclid \citep{euclidoverview2024} and LSST \citep{LSSToverview2019}, we are close to getting a statistically robust measurement of $\sigma_8$. It it thus imperative for models assuming $\Lambda$CDM to have high accuracy and precision, while keeping the door open to potential extensions.

The tension in $\sigma_8$ is currently at the $2-3\sigma$ level \citep{KIDS2021,DESYR3,HSC2023,McCarthy2023}. Surveys like Euclid and LSST will provide us with much tighter constraints based on weak lensing and galaxy clustering. Another promising avenue is to approach the tension from many different directions. In additions to lensing, we can, for example, use the Sunyaev-Zel'dovich (SZ) effect power spectrum \citep[e.g.][]{SZ1972,PlanckSZmap2014,SPTmap2022}, cluster counts \citep[e.g.][]{PlanckClustercosmo2016,Bocquet2024,Ghirardini2024}, CMB-lensing \citep[e.g.][]{PlanckLensing2020} and all their cross-correlations to constrain cosmology. These independent probes are not only a way to increase our constraints on $\sigma_8$, but, if the tensions persist, they can also provide valuable insight into how to extend the $\Lambda$CDM model.

Here we will investigate cosmological constraints via cluster count cosmology \citep[for reviews see][]{ClusterReviewAll2011,Prattreview2019}. The number of haloes as a function of halo mass is described by the halo mass function (HMF), which is sensitive to the cosmological parameters. In particular, the high mass end, where clusters are found, is very sensitive to changes in $\sigma_8$. Because of this sensitivity, even a statistic as simple as counting the number of observed clusters per unit volume can provide valuable constraints on $\sigma_8$ and, to a lesser extent, $\Omega_{\rm m}$, the matter content of the Universe.

In order to count clusters above some mass limit, we need an observable proxy for halo mass with which we can select clusters. As clusters are very massive objects, they can be observed across almost the entire electromagnetic spectrum. The three most favoured methods use the X-ray emission from the hot intra-cluster medium \citep[e.g.][]{XXL2018,eFEDScat2022,Eroscat2024}, the number of member galaxies in the optical \citep[e.g.][]{Reddragon2022,Euclidclusters2022}, and the distortion of the CMB spectrum due to their high electron pressures via the SZ effect \citep[e.g.][]{PLanckSZ2016,Hilton2018,Bleem2024}. Additionally, methods using the cluster lensing signal are starting to be used \citep{Hamana2020,Chen2024Lensclust}. \cite{Kugel2024} use the FLAMINGO cosmological simulations \citep{FLAMINGOmain,Kugel2023} to compare the performance of the different observables and found that SZ-selection is generally less prone to selection effects than X-ray or galaxy richness selection (using a window of size $5R_{\rm 500c}$\footnote{$R_{\rm 500c}$ is the radius around a halo within which the enclosed density is 500 times the critical density. The mass within this radius is defined as $M_{\rm 500c}$.}, as appropriate for the Planck satellite).

In order to measure cosmological parameters from cluster counts, we need accurate predictions for the expected number of clusters. The standard approach combines a theoretical HMF with a scaling relation between halo mass and an observable mass proxy with an assumed level of scatter around it. The scaling relation is further assumed to be a power-law with lognormal scatter with a fixed $\sigma_{\log_{10}}$ across the mass range \citep[e.g.][]{Rozo2014,Evrard2014,PlanckClustercosmo2016,SPTclustercosmo2019,SDSSClustercosmo2019,eFEDSclustercosmo2023,Ghirardini2024}. HMF models typically do not account for baryonic effects or do so assuming a particular galaxy formation model. These assumptions might induce biases in the analysis because it is unlikely that scaling relations are perfect power-laws with mass-independent and lognormal scatter \citep{Kugel2024} and because poorly constrained astrophysical processes associated with galaxy formation are expected to modify the halo masses \citep[e.g.][]{Velliscig2014,Cui2014,Cusworth2014,Bocquet2016,FLAMINGOmain} and to bias the total masses measured using weak gravitational lensing \citep{Debackere2021}. Getting a good grip on these assumptions and the associated systematic uncertainties is key in making sure we can derive unbiased cosmology results \citep{Angulo2012,Mantz2019}.

Compared with cosmic shear and galaxy clustering measurements, cluster counts are less constraining for cosmological parameters. However, thanks to ongoing and upcoming X-ray \citep[e.g.][]{Eroscat2024}, optical \citep[e.g.][]{Euclidclusters2022} and SZ surveys \citep[e.g.][]{Hilton2018,SimonsObs2019,Klein2024,Bleem2024} it is likely that the constraints will tighten in the near future. The main goal of this work is to investigate if and how standard models and assumptions used for cluster cosmology affect the cosmological inference from cluster counts.

The models of the HMF are usually based on cosmological dark matter only (DMO) simulations. One of the difficulties here is that to accurately predict galaxy clusters, simulations with a very large box size are required. Clusters are very rare, and a large enough sample is needed to make statistically robust predictions. In order to model the HMF different methods are used. A common method is the one described by \cite{Jenkins2001}, where an empirical formula is fitted to DMO simulations and is then used to predict the HMF directly from the matter power spectrum. This is the method employed by for example \cite{Tinker2008,Tinker2010}, \cite{Bocquet2016} and \citet{Euclid2023}. Additionally, emulators have also started to be used for the HMF \citep[e.g.][]{MiraTitanHMF2020}. In this case the HMF is obtained by interpolating  between a set of DMO training simulations of different cosmologies.

These standard models are mostly based on DMO simulations. However, it is known that the HMF is modified by baryonic effects. The effects are strongest for clusters with mass $M_{\rm 500c}\lesssim10^{14}\rm{M}_{\odot}$ \citep[e.g.][]{Velliscig2014,Bocquet2016,FLAMINGOmain}. In 
 addition to a DMO HMF, \citet{Bocquet2016} use the HMF obtained from the Magneticum hydrodynamical simulation \citep{Magneticum2014} to constrain the fitting parameters directly to the HMF predicted by the hydro simulation. If baryonic effects are not taken into account correctly, the analysis is inconsistent, as the scaling relations are based on the true halo masses measured using X-rays or weak lensing rather than from DMO simulations \citep[e.g.][]{PlanckSZ2014,Lovisari2015,Akino2022}. One (model-dependent) way to get around this inconsistency is to match lensing masses from hydrodynamical simulations to DMO halo masses \citep{Grandis2021b}. Using this method, the baryonic effects can be added via mass - Compton-Y scaling relation \citep{eFEDSclustercosmo2023,Bocquet2024,Ghirardini2024}

Baryons need to be considered to predict observable mass proxies other than lensing. Feedback processes associated with galaxy formation change the distribution of the gas, which will change the halo mass function and modify the functional form of, and the scatter in the mass – observable relation. Hydrodynamic simulations make direct predictions for both the number of clusters and the observable mass proxies. However, most state-of-the-art simulations, such as EAGLE \citep{Eagle2015} and IllustrisTNG \citep{TNG2018}, do not have volumes large enough to provide a statistical sample of clusters. For cluster counts, doing zooms of individual objects \citep[e.g.][]{Ceagle2017} is not feasible as we need volume-limited statistics. Although projects like The Three Hundred Project \citep{Threehundred2018} and TNG-cluster \citep{ClusterTNG2023} can use zooms to simulate a hundreds of clusters from a single volume, they are unable to provide volume-limited samples of clusters. For cluster counts it is necessary to use large volumes even if this means lowering the resolution relative to simulations focusing on galaxy evolution. This is the approach taken in projects like Cosmo-OWLS \citep{CosmoOwls2014}, BAHAMAS \citep{Bahamas2017} and MilleniumTNG \citep{MTNG2022}. In addition, it is necessary to vary the uncertain strength of feedback processes associated with galactic winds driven by star formation and particularly by Active Galactic Nuclei (AGN) and to do so in a manner constrained by observations relevant for cluster cosmology \citep{CosmoOwls2014,Bahamas2017}.

For this work, we will make use of the cosmological hydrodynamic simulations of the FLAMINGO project \citep{FLAMINGOmain,Kugel2023}. FLAMINGO is a suite of very large volume simulations designed specifically to investigate the interplay between effects due to baryonic processes, massive neutrinos, and cosmology. The simulation suite includes a hydrodynamic simulation in a $(2.8~\rm{Gpc})^3$ volume using $(5040)^3$ gas particles and many model variations in $(1.0~\rm{Gpc})^3$ volumes. The largest volume, contains 461 (4100) clusters of mass $M_{\rm 500c}> 10^{15}~\rm{M}_{\odot}$ ($5\times10^{14}~\rm{M}_{\odot}$) at $z=0$. The FLAMINGO variations include variations in feedback, cosmology and resolution. Due to its large volumes, FLAMINGO provides statistically significant samples up to very high halo masses. The subgrid stellar and AGN feedback in the simulation have been calibrated to match the observed low-redshift galaxy stellar mass function and cluster gas fractions. The simulation have been shown to provide provide a good match to X-ray observations of clusters at the profile level \citep{Braspenning2023}. FLAMINGO was calibrated using machine learning assisted Gaussian process emulation \citep{Kugel2023}, and the feedback variations can each be related to systematic shifts in the cluster gas fractions and/or the galaxy mass function. In addition, the suite includes simulations that use an alternative AGN feedback prescription that uses kinetic jets instead of thermally driven winds, which is calibrated to match the same observations as the thermal models.

For this work we assume that the predictions from the FLAMINGO simulation are the ground truth that we compare our models against. In \citet{Kugel2024} we found that, in FLAMINGO, the scatter around the mass-observable scaling relation for X-ray luminosity within $R_{\rm 500c}$ and for Compton-Y within $5R_{\rm 500c}$ has power-law tails. However, we also showed that for Compton-Y within $R_{\rm 500c}$, which is the aperture we use here, the scatter is close to lognormal, but the scatter still changes with mass and redshift. Additionally, the scaling relations deviate from single power-laws and change between different feedback variations \citep[][Fig.~15, Fig.~A1 respectively]{FLAMINGOmain,Braspenning2023}. We will investigate how deviations from assumed scaling relations and scatter influence cluster counts and the cosmology inferred from them. We choose to focus on SZ-selected samples since those are intrinsically the least biased compared to a mass-selected sample \citep{Kugel2024}. The systematic errors that we report here are thus likely smaller than for other selection methods. In this work we do not attempt to forward model the selection effects using virtual observations. In the observations, SZ-selection is accomplished by applying a matched filter to CMB maps at different frequencies \citep[see e.g.][]{Melin2006,Melin2012,Hilton2018}. At the higher resolutions of current and upcoming CMB surveys, this will introduce additional selection effects via, for example, source confusion, foregrounds \citep[see e.g.][]{Melin2018,Zubeldia2023}, and the effect of beam smearing. In this work we neglect these effects and leave them for a future study, noting that the data products needed for such a work are available within the FLAMINGO suite of simulations.

The paper is structured as follows. In Section~\ref{sec:methods} we describe the FLAMINGO simulations, the models and functional forms used to predict cluster counts, and how we use those predictions to constrain cosmological parameters. In Section~\ref{sec:results} we show how different assumptions and changes due to the variations in the FLAMINGO simulation suite affect the predictions for cluster counts and the inferred cosmology. We summarise the results in Section~\ref{sec:conc}.

\section{Methods}\label{sec:methods}
In this section we describe the methods we use to predict cluster counts using the FLAMINGO simulations. We briefly introduce the FLAMINGO simulation and how we identify haloes in \S\ref{sec:flamingo}. We describe our predictive model for cluster counts in \S\ref{sec:cluster_count_model}. The three sample definitions used throughout this paper are introduced in \S\ref{sec:samp_defs}. We finish this section by describing the likelihood and how it is sampled in \S\ref{sec:likeli}.

\begin{table*}
    \centering
    \caption{The model ingredients used for each line shown in the figures of this work. In each row the assumptions that deviate from the fiducial model are bold faced. In each figure the bold faced entries are used as the labels for the lines. The first column shows the four groups of assumption tested. The second column shows which figures the lines appear in. The third column shows the DMO HMF used. The fourth columns shows the assumed baryonic modification of the DMO HMF, where the entry "DMO" indicates no baryonic effects are accounted for. The fifth and sixth columns list, respectively, the median observable-mass scaling relation and the scatter about that relation. The final column indicates which case, explained in Section~\ref{sec:scalscattheory}, is used to compute the scaling relation and scatter related quantities. Entries that contain the name of one of the FLAMINGO variations indicate that that assumption is taken from that particular simulation.}
\begin{tabular}{|l|l|l|l|l|l|l|}
\hline
Assumption & Figure & DMO HMF & Baryonic HMF & $M_{\rm 500c}-Y_{\rm 500c}$ & Scatter & Case \\
\hline
Fiducial & All & L5p6\_m10\_DMO & DMO & L1\_m9 & L1\_m9 & (i) \\ \hline
DMO HMF & \ref{fig:HMF_comp},\ref{fig:big_rat_fig} & \textbf{Tinker (2010)} & DMO & L1\_m9 & L1\_m9 & (i) \\
 & \ref{fig:HMF_comp},\ref{fig:big_rat_fig} & \textbf{Bocquet (2016)} & DMO & L1\_m9 & L1\_m9 & (i)\\
 & \ref{fig:HMF_comp},\ref{fig:big_rat_fig} & \textbf{MiraTitanEmulator} & DMO & L1\_m9 & L1\_m9 & (i) \\ \hline

Baryon effect on the HMF & \ref{fig:HMF_bar},\ref{fig:big_rat_fig} & L5p6\_m10\_DMO & \textbf{L1\_m9} & L1\_m9 & L1\_m9 & (i) \\
& \ref{fig:HMF_bar},\ref{fig:big_rat_fig},\ref{fig:cornerplot} & L5p6\_m10\_DMO & \textbf{fgas+2$\boldsymbol{\sigma}$} & L1\_m9 & L1\_m9 & (i)  \\
& \ref{fig:HMF_bar},\ref{fig:big_rat_fig} & L5p6\_m10\_DMO & \textbf{fgas-2$\boldsymbol{\sigma}$} & L1\_m9 & L1\_m9 & (i)  \\
& \ref{fig:HMF_bar},\ref{fig:big_rat_fig} & L5p6\_m10\_DMO & \textbf{fgas-4$\boldsymbol{\sigma}$} & L1\_m9 & L1\_m9 & (i)  \\
& \ref{fig:HMF_bar},\ref{fig:big_rat_fig} & L5p6\_m10\_DMO & \textbf{fgas-8$\boldsymbol{\sigma}$} & L1\_m9 & L1\_m9 & (i)  \\
& \ref{fig:HMF_bar},\ref{fig:big_rat_fig} & L5p6\_m10\_DMO & \textbf{Jet} & L1\_m9 & L1\_m9 & (i) \\
& \ref{fig:HMF_bar},\ref{fig:big_rat_fig} & L5p6\_m10\_DMO & \textbf{Jet\_fgas-4$\boldsymbol{\sigma}$} & L1\_m9 & L1\_m9 & (i) \\ \hline

Scaling relation fit  & \ref{fig:PL_comp},\ref{fig:big_rat_fig} & L5p6\_m10\_DMO & DMO &  L1\_m9 & \textbf{LN}  & (iv) \\
 & \ref{fig:PL_comp},\ref{fig:big_rat_fig} & L5p6\_m10\_DMO & DMO & \textbf{PL} &  L1\_m9 & (iii) \\
 & \ref{fig:PL_comp},\ref{fig:PL_uncertainties},\ref{fig:big_rat_fig} & L5p6\_m10\_DMO & DMO & \textbf{PL} & \textbf{LN} & (ii)\\
 & \ref{fig:PL_comp},\ref{fig:big_rat_fig} & \textbf{Bocquet (2016)} & DMO & \textbf{PL} & \textbf{LN} & (ii) \\ \hline

Scaling relation variations & \ref{fig:scale_vars},\ref{fig:big_rat_fig} & L5p6\_m10\_DMO & DMO & \textbf{fgas+2$\boldsymbol{\sigma}$} & \textbf{fgas+2$\boldsymbol{\sigma}$} & (i) \\
 & \ref{fig:scale_vars},\ref{fig:big_rat_fig} & L5p6\_m10\_DMO & DMO & \textbf{fgas-2$\boldsymbol{\sigma}$} & \textbf{fgas-2$\boldsymbol{\sigma}$} & (i) \\
 & \ref{fig:scale_vars},\ref{fig:big_rat_fig} & L5p6\_m10\_DMO & DMO & \textbf{fgas-4$\boldsymbol{\sigma}$} & \textbf{fgas-4$\boldsymbol{\sigma}$} & (i) \\
 & \ref{fig:scale_vars},\ref{fig:big_rat_fig} & L5p6\_m10\_DMO & DMO & \textbf{fgas-8$\boldsymbol{\sigma}$} & \textbf{fgas-8$\boldsymbol{\sigma}$} & (i) \\
 & \ref{fig:scale_vars},\ref{fig:big_rat_fig} & L5p6\_m10\_DMO & DMO & \textbf{Jet} & \textbf{Jet} & (i) \\
 & \ref{fig:scale_vars},\ref{fig:big_rat_fig} & L5p6\_m10\_DMO & DMO & \textbf{Jet\_fgas-4$\boldsymbol{\sigma}$} & \textbf{Jet\_fgas-4$\boldsymbol{\sigma}$} & (i) \\  \hline
Full hydrodynamic variations & \ref{fig:cornerplot} & L5p6\_m10\_DMO & \textbf{fgas-8$\boldsymbol{\sigma}$} & \textbf{fgas-8$\boldsymbol{\sigma}$} & \textbf{fgas-8$\boldsymbol{\sigma}$} & (i) \\
 & \ref{fig:cornerplot} & L5p6\_m10\_DMO & \textbf{Jet} & \textbf{Jet} & \textbf{Jet} & (i) \\  \hline

\end{tabular}

\label{tab:ingre_sum}
\end{table*}

\subsection{FLAMINGO}\label{sec:flamingo}
To construct virtual cluster catalogues, we make use of the HMF, observable-mass scaling relations and their scatter obtained from the FLAMINGO simulations \citep{FLAMINGOmain,Kugel2023}. We will make use of the simulations at FLAMINGO's intermediate resolution ($m_{\rm gas}=1.09\times10^{9}~\rm{M}_{\odot}$) in box sizes of $(1~\rm{Gpc})^3$. These simulations use $2\times1800^3$ gas and dark matter particles, $1000^3$ neutrino particles, and all assume the Dark Energy Survey year 3 \citep{DESYR3} cosmology $(\Omega_{\rm m} = 0.306, \ \Omega_{\rm b} = 0.0486, \ \sigma_8 = 0.807, \ {\rm H}_{0} = 68.1~\rm{km/s/Mpc}, \ n_{\rm s} = 0.967)$. The simulations are run with the cosmological smooth particle hydrodynamics and gravity code \texttt{SWIFT} \citep{SWIFT2023} using the \textsc{SPHENIX} SPH scheme \citep{Sphenix2022}. The initial conditions are obtained from a modified version of \textsc{monofonIC} \citep{Monofonic2021,ElbersIcs2022}, and neutrinos are implemented with the $\delta f$ method \citep{Elbers2021}.

The FLAMINGO subgrid physics is an evolution the models developed for OWLS \citep{OWLS2010} and used in BAHAMAS \citep{Bahamas2017}. It includes radiative cooling \citep{Ploeckinger2020}, star formation \citep{SchayeDV2008}, stellar mass loss \citep{Wiersma2009chemistry}, kinetic supernova feedback \citep{DVSchaye2008kin,Chaikin2022,Chaikin2022b}, super massive black holes \citep{Springel2005a,bahe2021} and AGN feedback in both thermal \citep{BoothSchaye2009} and kinetic form \citep{Husko2022}. 

One of the new features is that the subgrid physics was calibrated by fitting the simulations to match the $z=0$ galaxy stellar mass function (in the mass range $10^{10}~\rm{M}_{\odot}\lesssim M_{*}\lesssim10^{11.5}~\rm{M}_{\odot}$) and the gas fractions in low-$z$ groups and clusters (up to a mass of $M_{\rm 500c}=10^{14.3}~\rm{M}_{\odot}$) using machine learning \citep{Kugel2023}. The same method was used to design a set of variations which are directly based on the observed error bars. Of particular interest for this work are the variations in the cluster gas fraction and AGN model, which are denoted as fgas$\pm N\sigma$ and Jet\_fgas$\pm N\sigma$. For these models the $N\sigma$ denotes by how many observed standard deviations the gas fractions have been shifted up or down with respect to the fiducial model, which is based directly on the observations. The Jet models make use of kinetic jets for the AGN feedback instead of the thermally-driven winds used for all other runs. These Jet models are calibrated to match the same data as the corresponding thermal AGN feedback models. When fitting the gas fraction for the fiducial model and variations, we fit over the mass range $M_{\rm 500c}=10^{13.5}~\rm{M}_{\odot}$ to $10^{14.36}~\rm{M}_{\odot}$. The calibration overlaps only slightly with the objects of the masses that are investigated in this paper, which for our deepest mock sample go down to $M_{\rm 500c}\approx 2\times 10^{14}~\rm{M}_{\odot}$ (see Table~\ref{tab:sample_defs}). The gas fractions for the different models differ most at $M_{\rm 500c}\sim 10^{14}~\rm{M}_{\odot}$, towards higher mass the differences between the feedback variations become smaller. This is both due to the fact that we do not calibrate above a mass of $M_{\rm 500c}=10^{14.36}~\rm{M}_{\odot}$ and because feedback is unable to offset the inflow of gas in the most massive systems as they end up being mostly baryonically closed. For $M_{\rm 500c}= 10^{14.5}~\rm{M}_{\odot}$ the median gas fractions increase from 0.090 for fgas$-8\sigma$ to 0.115 for fgas$+2\sigma$. At $M_{\rm 500c}= 10^{15}~\rm{M}_{\odot}$ the models span gas fractions between 0.113 and 0.122. More details can be found in \citet{FLAMINGOmain}, Fig.~10 of which shows the median gas fraction as a function of mass for all models.

To obtain catalogues of (sub)haloes from the simulation, we make use of a modified version of Hierarchical Bound Tracing+ (HBT+) \citep[][Forouhar Moreno et al. in prep]{HBT2012,HBT2017}. The modifications include extensions required for the application to hydrodynamic simulations. HBT+ tracks haloes through "history space", as it uses the past halo membership of individual particles to help identify where their host haloes are located at later times. This approach is possible thanks to hierarchical structure formation, as any present-day (satellite or central) halo was a central when it first formed. This means that particles are associated to field haloes when they form, which are easy to locate using a 3D Friends-of-Friends (FOF) algorithm. The association of particles to a given halo is saved for subsequent output times. This information is then used to identify where the haloes are located at subsequent times, and results in more robust identification of satellite haloes. This also means that past information can help identify which halo is the central within a FOF group, and reduce the problem of "central" swapping.

For each central halo in the HBT+ catalogue we use the Spherical Overdensity and Aperture Processor (SOAP\footnote{\href{https://github.com/SWIFTSIM/SOAP}{https://github.com/SWIFTSIM/SOAP}}, McGibbon et al. in prep) to calculate $M_{\rm 500c}$ and the integrated Compton-Y. The Compton-Y contribution of the particle with index $i$, $y_i$, is stored in the snapshots and is calculated via
\begin{equation}
    y_i d_\text{A}^2 = \frac{\sigma_{\rm T}}{m_{\rm e} c^2} n_{\text{e},i} k_{\rm B} T_{\text{e},i} \frac{m_i}{\rho{}_i},
\end{equation}
where $d_\text{A}$ is the angular diameter distance, $\sigma_{\rm T}$ is the Thomson cross section, $m_{\rm e}$ is the electron mass, $c$ is the speed of light, $k_{\rm B}$ is the Boltzmann constant,  $n_{\text{e},i}$ is the electron number density, $T_{\text{e},i}$ is the electron temperature, $m_i$ is the mass and $\rho_i$ is the density of the particle. The contribution for each halo can be found by summing the contribution of all particles within an aperture. In this work we use the integrated Compton-Y within $R_{\rm 500c}$, i.e.\ we sum the quantity $y_i$ over all particles within $R_{\rm 500c}$.

\subsection{The cluster count model}\label{sec:cluster_count_model}
There are five key ingredients to predict the number of clusters that are observed in a cluster survey:
\begin{enumerate}
    \item The number of haloes per unit mass and volume via the HMF.
    \item The effect of baryonic physics on the HMF.
    \item The scaling relation between the median mass of the halo as a function of the observable.
    \item The scatter around the assumed observable-mass scaling relation.
    \item The probability for a halo with a certain observable property to be observed in the survey via the selection function.
\end{enumerate}
In order to always make our choices clear for each of these ingredients, we have summarised our choices used for each line in the paper in Table~\ref{tab:ingre_sum} (i-iv) and Table~\ref{tab:sample_defs} (v). Our primary goal is to investigate the effect of the different assumptions that are made for the first two of these items. Because we focus on the deviations on the theory side, we omit from our analysis the conversion between the observable and the signal measured at the telescope. Using these assumptions, we can predict the number of clusters using the following integral
\begin{align} \label{eq:number_count_integral}
    N(X_{\rm C}) = & \int_0^{z_{\text{max}}}\int_{0}^{A_{\text{sky}}}\int_{M_{\rm 500c,min}}^{M_{\rm 500c,max}}\phi(M_{\rm 500c},z,\boldsymbol{\theta},a_{\rm Astro}) ~\times \nonumber \\ & \chi(M_{\text{500c}},z,X_{\text{C}})\frac{\text{d} V}{\text{d}\Omega\text{d}z}(\boldsymbol{\theta})\text{d}M_\text{500c}\text{d}\Omega\text{d}z,
\end{align}
where $\phi(M_{\rm 500c}, \boldsymbol{\theta}, a_{\rm Astro},z)$ is the HMF, which depends explicitly on the astrophysical parameters $a_{\rm Astro}$, like the gas fraction on clusters and the assumed AGN model, $\chi(M_{\rm 500c},z,X_{\rm C})$ gives the probability of observing a cluster with mass $M_{\rm 500c}$ at redshift $z$ given a cut (i.e. selection limit) $X_{\rm C}$ on observable $X$, $A_{\text{sky}}$ is the survey angular area on the sky, and $\frac{\text{d} V}{\text{d}\Omega\text{d}z}(\boldsymbol{\theta})$ is the differential comoving volume, which, like the HMF, depends on the cosmological parameters $\boldsymbol{\theta}$. We note here that in the integral $M_{\rm 500c}$ refers to the true total mass, including baryonic effects. We integrate from redshift zero to $z_{\rm max}=2$, noting that going to higher redshifts makes a negligible difference for the cuts we are interested in. For $M_{\rm 500c}$ we integrate from $10^{12}~\rm{M}_{\odot}$ to $10^{16}~\rm{M}_{\odot}$, noting that we run out of haloes before we reach the maximum mass. The minimum mass of $10^{12}~\rm{M}_{\odot}$ ensures we do not miss any haloes for the observational selection cuts we apply. We are specifically interested in selecting clusters via their Compton-Y contribution within $R_{\rm 500c}$. In this case $X$ becomes $Y_{\rm{500 c}}$, which we will denote simply as $Y$, with the cut being defined as $Y_{\rm C}$.
\subsubsection{The selection function}
For the selection function we use a simple step function
\begin{equation}
    P(Y|Y_{\rm C}) = \begin{cases}
    1 &\text{ if }Y>Y_{\rm C}, \\
    0 &\text{ if }Y<Y_{\rm C}. \\
    \end{cases}
\end{equation}
Since we are interested in the effect of theoretical uncertainties, this selection function gives us results that are easily interpretable. The functional form of the selection function is identical for all mock samples in this work. To investigate the required level of accuracy for different mock samples, we will vary the selection cut until the total number of objects in the mock sample roughly matches to a few observational samples. This will be discussed in more detail in Section~\ref{sec:samp_defs}.

\subsubsection{The observable-mass scaling relation and scatter}\label{sec:scalscattheory}

To obtain $\chi(M_{\rm 500c},z,X_{\rm C})$, we use $P(Y|\hat{Y}(M_{\rm 500c},z))$. Here $\hat{Y}$ is the median value of $Y$ at a given $M_{\rm 500c}$ and $z$ as given by the observable-mass scaling relation. We consider four cases designed to separate the effects of two of the different model ingredients: the power-law observable-mass scaling relation and the associated scatter:
\begin{enumerate}
    \item Taking both the scaling relation and scatter from the hydrodynamical simulation.
    \item Using a power-law scaling relation with lognormal scatter, both fit to the simulation results, referred to as "PL+LN".
    \item Using the scaling relation from the simulation with lognormal scatter fit to the simulation results, referred to as "LN".
    \item Using a power-law scaling relation fit to the simulation results with scatter from the simulation, referred to as "PL".
\end{enumerate}
By comparing the four cases we can assess potential systematics due to conventional assumptions. As our main interest is comparisons with FLAMINGO, our fiducial cluster count model will use case (i). We refer the reader to Fig.~15 of \citet{FLAMINGOmain} for the FLAMINGO Compton-Y - $M_{\rm 500c}$ scaling relation. The scatter about this relation is shown in detail in Figures~1 and B1 of \citet{Kugel2024}. When describing the changes between these four cases, it is important to remember that the quantity we need is $\chi(M_{\rm 500c},z,X_{\rm C})$, the probability of observing an object with a given mass and redshift given a selection cut. 

\begin{figure}
    \centering
    \includegraphics[width=.99\columnwidth]{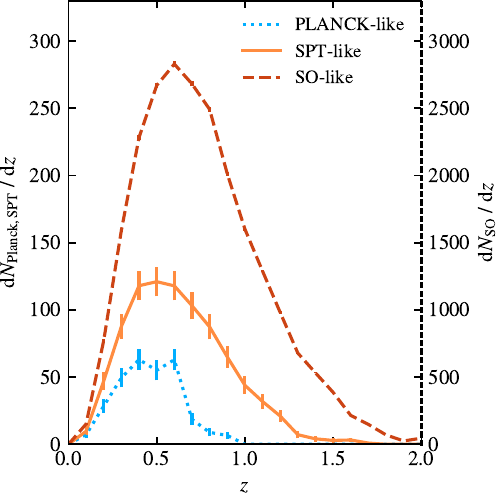}
    \caption{The redshift distribution of cluster number counts for our fiducial model (L5p6\_m10\_DMO\ HMF, $Y-M_{\rm 500c}$ scaling relation and scatter taken from simulation L1\_m9). The three different lines show the predictions for the three mock sample sizes we investigate in this paper. The error bars show the Poisson errors. Note that the PLANCK- and SPT-like surveys use the left y-axis, while the SO-like survey uses the right y-axis. Note also that the SO-like errors are smaller than the line width.}
    \label{fig:diff_surveys}
\end{figure}

In case (i), where we use both the scaling relation and the scatter from the hydrodynamical simulation, we use only the binned outputs of the simulation. To obtain the scaling relation from the simulations, we set up a grid in $M_{\rm 500c}$ and $z$. We sample the mass in bins of width 0.1 dex between $M_{\rm 500c} = 10^{12}~\rm{M}_{\odot}$ and $10^{16}~\rm{M}_{\odot}$, and we sample the redshift in bins of $0.05$ from $z=0$ to $z=2$, which is the redshift sampling of snapshots for FLAMINGO. In each bin we calculate the number density of objects and the median value for $Y$ using the snapshot of the simulation at the given redshift. Additionally, for each mass-redshift bin we use 800 bins between a $Y$ of $10^{-10}$ and $10^{-2}$ to sample the distribution of $Y$. As the quantities of interest are statistical in nature, we use the entire snapshot volume at each redshift to construct these grids. To obtain the probability of observing a halo in this mass-redshift bin, we want to find the fraction of haloes that have $Y>Y_{\rm C}$. For our selection function, this is equivalent to
\begin{align}
    \chi(M_{\rm 500c},z,X_{\rm C}) &= \int_{-\infty}^{\infty}P(Y|\hat{Y}(M_{\rm 500c},z))P(Y|Y_{\rm C}){\rm d}Y , \nonumber \\
    &= 1-C_{M_{\rm 500c},z}(Y_{\rm C}) ,
\end{align}
where $C_{M_{\rm 500c},z}(Y_{\rm C})$ is the normalised cumulative distribution function (CDF) at fixed mass and redshift. For each mass-redshift bin we compute the CDF by interpolating the 800 $Y$ bins, giving us a smooth function. The interpolation is required to be able to convert between the FLAMINGO median, and the parametric fitted median. As the simulation binning is already a combination of the scatter and scaling relation, we do not need to take any additional steps when the goal is using both the scaling relation and scatter from the simulations. We use a large number of bins as we only use the bins for the CDF, allowing for high precision spacing while conserving the smooth behaviour. When the mass-redshift bin of the scaling relation is empty, the bin does not contribute to the cluster counts and it is set to be unobservable, even if the HMF does predict haloes in that bin. We have verified that this leads to deviations that are far smaller than the Poisson errors.

In case (ii) of using a power-law scaling relation with lognormal scatter, we instead assume a functional form to obtain the required probabilities. The functional form we use was used by \citet{PlanckSZ2014,PlanckClustercosmo2016} and \citet{Salvati2022} and is given by
\begin{equation}
    E^{-\beta}(z)\left[\frac{D^2_{\rm A}(z)\hat{Y}}{10^{-4}~\rm{Mpc^2}}\right]=Y_{*}\left[\frac{h}{0.7}\right]^{-2+\alpha}\left[\frac{(1-b)M_{\rm 500c}}{6\times10^{14}~\rm{M}_{\odot}}\right]^{\alpha},\label{eq:powerlaw}
\end{equation}
where the fitting parameters $Y_{*}$, the power-law amplitude, $\alpha$, the power-law slope, $\beta$, the slope of the redshift scaling, and $b$, the hydrostatic bias, are fit to the results from the simulation. Any change is thus purely due to slight deviations from a power-law in the simulation. The values and priors can be found in Table~\ref{tab:params}.

\begin{table}
    \centering
    \caption{Priors on the cosmological and power-law fit parameters (Eq.~\ref{eq:powerlaw}) taken from the MiraTitanEmulator and \citet{PlanckClustercosmo2016} respectively and the fiducial FLAMINGO values for each parameter. Here $\mathcal{U}(x,y)$ stands for a uniform distribution between $x$ and $y$, $\mathcal{N}(\mu,\sigma)$ stands for a normal distribution with mean $\mu$ and standard deviation $\sigma$. The last two parameters, $\beta$ and $b$, are fixed, though we do give the uncertainty used by \citet{PlanckClustercosmo2016} on  the parameters  for context.}
    \begin{tabular}{l|l|l}
    \hline Parameter  &  Prior & FLAMINGO\\ \hline
    $\Omega_{\rm m}$  & $\mathcal{U}(0.259,0.334)$ & 0.306\\
    $\sigma_{8}$ & $\mathcal{U}(0.7,0.9)$ & 0.807\\ \hline
    $\log_{10}Y_{*}$ & $\mathcal{N}(-0.19,0.02)$ & $-0.098$\\
    $\alpha$ & $\mathcal{N}(1.79,0.08)$ & 1.66\\
    $\sigma_{\log_{10}}$ & $\mathcal{N}(0.075,0.01)$ & 0.081\\
    $\beta$ & 0.66~($\pm 0.50$) & 0.89\\
    $b$ &  $\mathcal{N}(0.780,0.092)$ & 0.743\\\hline
    \end{tabular}
    \label{tab:params}
\end{table}
For lognormal scatter we can compute the CDF via
\begin{equation}\label{eq:lognrmscat}
    1-C_{M_{\rm 500c},z}(Y_{\rm C}) = \frac{1}{2}\left[1-\text{erf}\left(\frac{\log_{10}Y_{\rm C} - \log_{10}\hat{Y}(M_{\rm 500c},z)}{\sqrt{2}\sigma_{\log_{10}}}\right)\right],
\end{equation}
where $\sigma_{\log_{10}}$ is the standard deviation parameterising the lognormal scatter in $Y$ in dex and $\hat{Y}(M_{\rm 500c},z)$ is the median. The prior on $\sigma_{\log_{10}}$ is taken from \cite{PlanckSZ2014,PlanckClustercosmo2016}. 

For case (iii), where we use the observable-mass scaling relation from the simulation but with lognormal scatter, we note that the CDF given by Eq.~\ref{eq:lognrmscat} has the scaling relation as a direct input. So we can use either the power-law from Eq.~\ref{eq:powerlaw} or directly use the simulation prediction, covering cases (ii) and (iii), respectively. We use the same level of scatter for both, i.e. for case (ii) we do not fit the scatter around the median but always use the fixed value obtained from the simulation. However, as this simulation scatter itself is very close to lognormal, fitting a separate value for both would likely have no significant effect on the results.

Our final case, case (iv), is using a power-law for the observable-mass scaling relation, with scatter from the simulation. In this case we again make use of the binned interpolators that are used for the simulation only case. To conserve the scatter, we calculate for each mass and redshift bin what the difference is between the power-law scaling relation fit to the simulation prediction and the actual simulation scaling relation. We then recenter the distribution in each mass-redshift bin on the power-law predictions. This conserves the shape of the scatter in the simulation, but the median now follows the power-law scaling relation.

The choices of scaling relation and scatter for each line shown in the figures in the rest of the paper are listed in the last three columns of Table~\ref{tab:ingre_sum}. This includes which of the four cases discussed here applies to each specific setup.

\subsubsection{The HMF}

To obtain the HMF from the simulation, we use the same binning approach as used for the scaling relation (0.1 dex in $M_{\rm 500c}$ and steps of $0.05$ in $z$) to get the number density of objects at each mass and redshift using a the entire volume of each snapshot of the simulation at each redshift. In addition to the simulation results, We investigate how well the DMO HMF models of \cite{Tinker2010}, \cite{Bocquet2016} and \citet[][which we will refer to as the MiraTitanEmulator]{MiraTitanHMF2020} agree with the DMO HMF of the FLAMINGO simulations used in this work. They can be divided into two categories. The models by \cite{Tinker2010} and \cite{Bocquet2016} are based on the fitting formula introduced by \cite{Jenkins2001} and they derive the HMF directly from the linear theory matter power spectrum. The MiraTitanEmulator is an emulator that was trained on the MiraTitan set of DMO simulations \citep{MiraTitan2016}. \citet{Tinker2010} and \citet{Bocquet2016} directly provide the HMF based on $M_{\rm 500c}$. For the MiraTitanEmulator, we convert from $M_{\rm 200c}$ to $M_{\rm 500c}$ assuming an NFW density profile and using the mass-concentration relation by \cite{Diemer2019}. The DMO HMF used for the different lines in the work is listed in column three of Table~\ref{tab:ingre_sum}. The cosmology of the different models is always fixed to the FLAMINGO cosmology to ensure fair comparisons. However, when we vary the cosmology in Section~\ref{sec:fitall}, we are unable to use the FLAMINGODMO HMFs as they are only for a fixed cosmology. Instead, we will make use of the \citet{Bocquet2016} DMO HMF.

All these HMFs model the DMO HMF. To investigate the effects of baryons on the HMF, we allow for the following correction
\begin{equation}
    \phi(M_{\rm 500c, Hydro},z) = \phi_{\rm DMO}(M_{\rm 500c, DMO},z) F_{\rm cor,bar}(M_{\rm 500c, DMO}, z),
\end{equation}
where $F_{\rm cor,bar}(M_{\rm 500c}, z)$ gives the correction factor to convert from a DMO HMF to the HMF affected by baryons. We measure the correction factor from the FLAMINGO simulations using the same binning for each HMF as described earlier. As we measure the ratio as a function of the corresponding DMO halo mass, this leads to a self-consistent conversion. If either the DMO or hydro HMF has no haloes in a bin, we set the ratio to one. We note here that not correcting the DMO HMF for baryonic effects causes an inconsistency in the calculation as the scaling relations are always based on the hydrodynamic masses. The baryonic correction used for each line is shown in column four of Table~\ref{tab:ingre_sum}.

\subsubsection{Evaluating the integral}

Having specified each of its individual ingredients, we can now evaluate the integral in equation~\ref{eq:number_count_integral} and predict the cluster counts. To ensure that the number counts vary continuously as a function of redshift, which is necessary as we want to be able to deviate from the simulation snapshot redshifts when defining our redshift bins, we first perform the following integral
\begin{align}
    \frac{\text{d}N(Y>Y_{\text{C}})}{\text{d}z}(z) = &\int_{0}^{A_{\text{sky}}}\int_{M_{\text{500c,min}}}^{M_{\text{500c,max}}}\phi(M_{\text{500c}},z,\boldsymbol{\theta},a_{\text{Astro}}) ~ \times \nonumber \\ & \chi(M_{\text{500c}},z,Y_{\text{C}})\text{d}M_{\text{ 500c}}\text{d}\Omega,
\end{align}
resulting in the redshift distribution of the number counts. We use linear interpolation to obtain a continuous version of the distribution. We then numerically integrate this continuous function for each bin using a quadratic integrator. We have verified that these choices do not lead to significant numerical errors.

As we have continuous analytic alternatives for each of the ingredients of the cluster count model, the DMO HMF models from the literature and the lognormal and power-law assumptions, we have verified, by evaluating the number counts using \texttt{quad} integration, that our choice of binning (0.1 dex in $M_{\rm 500c}$ and steps of $0.05$ in $z$) does not lead to significant numerical errors. Therefore, we evaluate our continuous assumptions on the same grid of inputs as the simulations. This allows for fair comparisons between the simulation and analytic predictions and allows for a significant speed up of the code.

\subsection{Definition of samples}\label{sec:samp_defs}
We will use three sample definitions, loosely based on available SZ cluster samples. We include a shallow full-sky sample, using $Y_{\rm C}=10^{-4}~\rm{Mpc}^{-2}$. This is predicted to have 360 clusters for our fiducial simulation, which is similar to the number of objects in the Planck survey \citep{PlanckClustercosmo2016}. We also include a $5000~\rm{deg}^2$ survey with a cut of $Y_{\rm C}=3\times10^{-5}~\rm{Mpc}^{-2}$. This predicts around 1000 objects, making it similar to the current SPT survey \citep{Bocquet2024}. As a reference for future observations, we also include a Simons Observatory (SO)-like survey. This survey covers 40\% of the sky with a cut of $Y_{\rm C}=10^{-5}~\rm{Mpc}^{-2}$. This predicts approximately 27000 objects, in line with SO forecasts \citep{SimonsObs2019}. The sample characteristics are summarised in Table~\ref{tab:sample_defs}. Our three samples are defined by their respective selection limits. As explained before, this choice is independent of any choices made for the model assumptions. Hence, for each line shown in Table~\ref{tab:ingre_sum} we can make a prediction for each of the three sample definitions. We base our mocks on observed samples that are cross-matched with optical surveys to obtain redshifts for each object, and we include the redshift information. Hence, to create model data, we sample the distribution at every $\Delta z= 0.1$ in redshift (i.e.\ a coarser redshift sampling than used for the simulations) between $z=0$ and $z=2$, resulting in $19$ bins. We want to stress again here that we base these samples on the statistical distributions given by either FLAMINGO or the models explained in the previous subsection. We do not construct virtual lightcones within the simulation.

\begin{table}
    \caption{For each of the three mock sample setups used throughout this work the table lists the SZ cut, fraction of the sky covered, and the approximate $M_\text{500c}$ mass cuts that the SZ cut corresponds to at $z=0$ and 1. To find the mass cut, we take the median mass at the SZ cut in the L1\_m9 simulation. Note that for Planck we do not have any objects that are above the mass cut at $z=1$.}
    \centering
    \begin{tabular}{l|r|r|r|r}
      \hline Sample  & $Y_{\rm C}$ [Mpc$^{-2}$] & $A_{\rm sky}$ & $M_{{\rm C},z=0}~[\rm M_{\odot}]$ & $M_{{\rm C},z=1}~[\rm M_{\odot}]$\\ \hline
       Planck  & $10^{-4}$ & 1.00 & $9.2\times10^{14}$ & - \\
       SPT & $5\times10^{-5}$ & 0.12 &$4.5\times10^{14}$ & $3.7\times10^{14}$\\
       SO & $10^{-5}$ & 0.40 & $2.3\times10^{14}$ & $1.9\times10^{14}$ \\ \hline
    \end{tabular}
    \label{tab:sample_defs}
\end{table}

We want to be able to compare all the different assumptions to a single "Fiducial" set of assumptions taken from the FLAMINGO simulations. For our fiducial cluster count model we will assume the observable-mass scaling relation from the hydro $(1~\rm{Gpc})^3$ volume, L1\_m9, combined with the DMO HMF from our biggest (5.6~Gpc)$^3$ DMO volume. As shown by \citet{FLAMINGOmain}, the DMO HMFs agree between the different resolutions, so we use the largest volume in order to reduce the noise. We thus use our calibrated model for the hydro, combined with a large volume to reduce cosmic variance and Poisson errors in the DMO HMF. This does imply that our fiducial model has the common inconsistency that the scaling relation is based on hydro masses, while the HMF uses DMO masses, without correcting for it. We choose to investigate this potential systematic effect separately. For our fiducial cluster count model, the distribution of sources with redshift is shown in Fig.~\ref{fig:diff_surveys}. From the figure it is clear that the relative error bars become much smaller for the larger samples. There is a also a slight shift towards sampling higher redshift objects with deeper surveys. Due to the limited redshift range of the MiraTitanEmulator, we are unable to use its predictions for number counts at $z>2$. However, from this figure it is clear that this leads to missing a negligible fraction of all sources, even for the deepest survey.

We want to make a small note here on the error bars, which are assumed to be Poissonian throughout this work. The Poisson errors are obtained for each bin independently, and depend only on the number of objects in each bin. This is the error that is representative of the best case scenario for the actual surveys. However, these are the errors on a virtual lightcone. For our fiducial cluster count model we sample the observable-mass scaling relation from a $(1~{\rm Gpc})^3$ volume, which has a cosmic variance that is thus separate from the Poisson errors shown here. For the SPT- and SO-like surveys this source of noise has no impact on our results, however, for the Planck-like samples, there are a few cases where this noise leads to problems due to the relatively small number of very massive clusters in the (1Gpc)$^3$ volume.
\begin{figure*}
    \centering
        \includegraphics[width=.99\columnwidth]{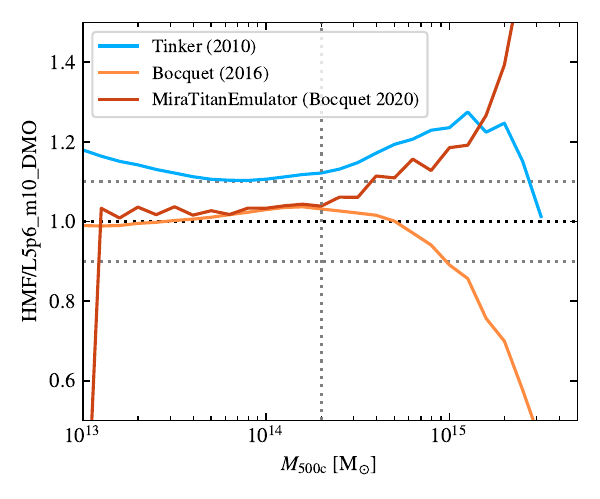}        \includegraphics[width=.99\columnwidth]{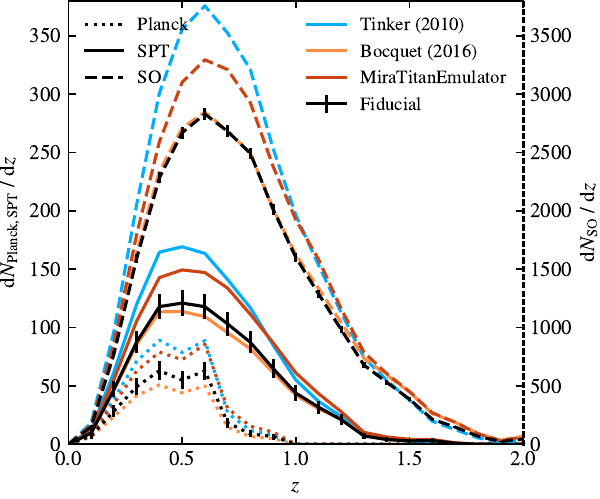}
    \caption{The left panel shows the ratio between the three HMF models specified in the legend and the FLAMINGO L5p6\_m10\_DMO for the FLAMINGO cosmology at $z=0$. The black dotted horizontal line shows the one-to-one line and the grey dotted horizontal line shows the ten per cent region. The grey dotted vertical line roughly coincides with the lowest cluster masses that will be in the deepest (SO) sample. The right panel shows the distribution of clusters with redshift for the three different surveys, given by different line-styles. The different coloured lines show the cluster counts when assuming different HMF models. The black line with error bars shows the cluster counts from the fiducial model (L1\_m9 observable-mass scaling relation and L5p6\_m10\_DMO HMF). The error bars indicate the Poisson errors in each bin. Note that the left y-axis is for the dotted (Planck) and solid (SPT) lines, while the right y-axis is for the dashed lines (SO).}
    \label{fig:HMF_comp}
\end{figure*}
\subsection{The likelihood}\label{sec:likeli}
To fit for cosmological parameters, we make use of a Poisson likelihood. While we use the snapshot outputs for all our predictions, we have separately verified that this is a valid assumption by bootstrap resampling different parts of the sky using FLAMINGO's halo lightcone output, which links haloes from the snapshots to their locations in the FLAMINGO particle lightcone outputs that were produced on-the-fly by using the ID of each halo's central black hole particle. Taking the model counts $\lambda_i$ and the observed counts $x_i$ for each redshift bin $i$, the likelihood of binned data $\mathbf{x}$ being the result of model data $\boldsymbol{\lambda}$ can be expressed as
\begin{align}
    P(\mathbf{x}|\boldsymbol{\lambda}) &= \Pi^N_i\frac{e^{-\lambda_i}\lambda_i^{x_i}}{x_i!},\\
    \ln P &=\sum^N_i \left(x_i\ln\lambda_i - \lambda_i - \ln(x_i!)\right),
\end{align}
where we will use $\ln P$ as the log likelihood. For our cosmological fits we vary only $\Omega_{\rm m}$ and $\sigma_8$, keeping all the other values fixed to the values from \cite{DESYR3}, the cosmology used for FLAMINGO. For the cosmological parameters we use flat priors that are based on the available parameter space for the MiraTitanEmulator, which can be found in Table~\ref{tab:params}. We use the publicly available package \texttt{emcee} \citep{emcee2013} using the ensemble sampler. We run the chains using 40 walkers for 2500 steps, where the first 500 are discarded.

To quantify the absolute and statistical biases of our posteriors with respect to the truth, we employ two different measures of the bias. The first gives us the fractional error of the median compared with the truth
\begin{equation}
    \text{Fractional error}=\frac{x-\mu}{\mu},
\end{equation}
where $x$ is the median and $\mu$ is the true value. We calculate the fractional error separately for $\sigma_8$ and $\Omega_{\rm m}$. Additionally, we calculate how many sigmas the truth is away from the posterior median values using the geometric distance assuming a covariant Gaussian
\begin{equation}
    N\sigma = \sqrt{(x-\mu)^{T}\Sigma^{-1}(x-\mu)},
\end{equation}
where $\Sigma$ is the covariance matrix. We provide these two different bias estimates to give an indication of the level of precision at which certain biases start to become dominant and to show how statistically significant the biases are for a purely cosmic-variance limited cosmological inference. Hence, the number of $\sigma$ should be interpreted as a worst case scenario for a given fractional error, as observational uncertainties will likely significantly reduce this number.

\section{Results}\label{sec:results}
In this section we will investigate how the following assumptions and uncertainties affect cluster counts:
\begin{itemize}
    \item The DMO HMF model (\S\ref{sec:hmf_comp})
    \item The effect of baryons on the HMF (\S\ref{sec:barhmf})
    \item The functional forms for the analytic observable-mass scaling relation and scatter (\S\ref{sec:models})
    \item The effect of changes in the assumed scaling relation (\S\ref{sec:scalvars})
\end{itemize}
For the effect of baryons on the HMF and the effect of changes in the assumed scaling relation we will make use of the variations in gas fraction and AGN feedback model in the FLAMINGO suite of simulations \citep{FLAMINGOmain,Kugel2023}. After investigating the effect on cluster counts for each type of assumption separately, we will compare the biases induced by different methods (\S\ref{sec:biasquant}) and see quantitatively how they impact the inference of $\sigma_8$ and $\Omega_{\rm m}$ (\S\ref{sec:fitall}). We refer the reader to Table~\ref{tab:ingre_sum} as a reference for the assumptions that are varied in the comparisons in the next sections.

\subsection{DMO halo mass function models}\label{sec:hmf_comp}
\begin{figure*}
    \centering
        \includegraphics[width=.99\columnwidth]{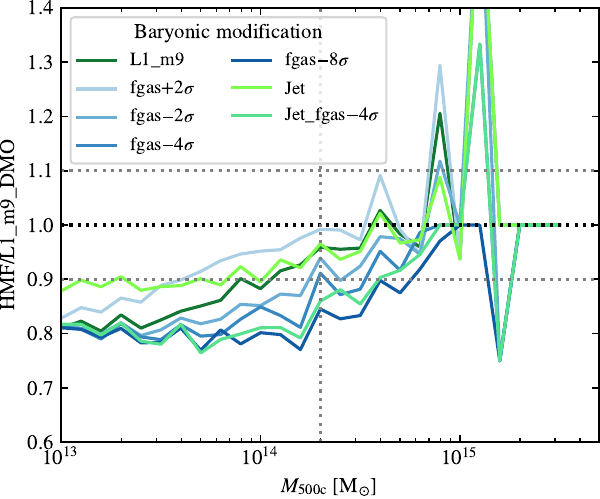}
        \hspace{0.5 cm}
        \includegraphics[width=.99\columnwidth]{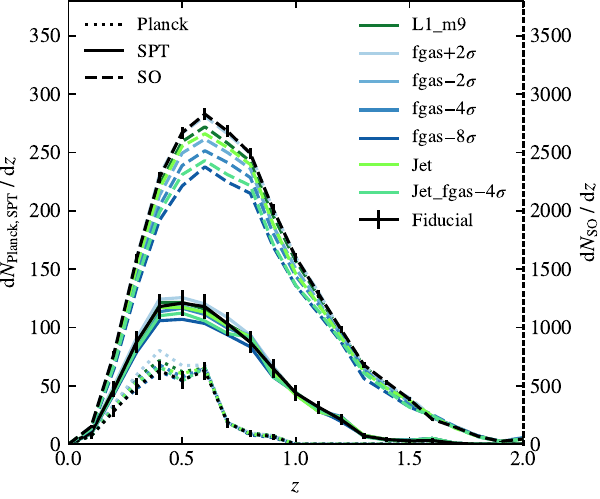}
    \caption{The left panel shows the ratio between the  HMF of the different FLAMINGO feedback variations, given by the legend, and the FLAMINGO L1\_m9\_DMO at $z=0$. The black dotted horizontal line shows the one-to-one line and the grey dotted horizontal lines show ten per cent deviations. The grey dotted vertical line roughly coincides with the lowest mass clusters that will be in the deepest (SO) sample. The functions in the left panel are used to modify the HMF of the fiducial DMO model and to create the right panel which shows the distribution of clusters with redshift for the three different surveys, indicated by different line-styles. The different coloured lines show the effect that baryons have on the HMF for each FLAMINGO simulation. The black lines with error bars shows the cluster counts from the fiducial model (L1\_m9 observable-mass scaling relation and L5p6\_m10\_DMO HMF). The error bars are the Poisson errors in each bin. Note that the left y-axis is for the dotted (Planck) and solid (SPT) lines, while the right y-axis is for the dashed lines (SO).}
    \label{fig:HMF_bar}
\end{figure*}
First we investigate the effect of different models for the DMO HMF. Fig. 19 of \citet{FLAMINGOmain} shows that the FLAMINGO DMO HMF agrees well with models taken from the literature. However, those plots were made using the Velociraptor \citep{VR2019} halo finder for $M_{\rm 200m}$, while we use the HBT+ \citep{HBT2017} halo finder and $M_{\rm 500c}$. Furthermore, some systematic deviations were found in \citet{FLAMINGOmain} that could lead to biases. We focus on comparing three widely used DMO HMF models at fixed cosmology with FLAMINGO: \cite{Tinker2010}, \cite{Bocquet2016} and the MiraTitanEmulator \citep{MiraTitanHMF2020}. We compare their predictions for the HMF and the resulting cluster counts for the FLAMINGO cosmology and compare with our fiducial setup, for which the HMF is based on the L5p6\_m10\_DMO simulation. We investigate the level of agreement between the models in Fig~\ref{fig:HMF_comp}. In the left panel we show the ratio between the different HMF models and the L5p6\_m10\_DMO FLAMINGO simulation at $z=0$. In the right panel we show the distribution of clusters with redshift for the three different mock samples indicated with different line styles. In both panels the colours indicate the different HMF models.

The \citet{Tinker2010} model and the MiraTitanEmulator fall outside of the error bars for all three mock samples. As can be seen in the left panel, the \cite{Tinker2010} HMF over-predicts the number counts at all masses. From the right panel it is clear that the over-prediction is worst at intermediate redshifts, but that the agreement becomes better for $z\geq1$. However, this is beyond the redshift range that contain most objects for our three mock samples. 

The MiraTitanEmulator is relatively accurate at low mass, overshooting by only a few per cent up to a mass of $M_{\rm 500c}\approx2\times10^{14}~\rm{M}_{\odot}$. Above this mass there is a large upturn and the model over-predicts the number of massive clusters. In the right panel it can be seen that this leads to a systematic over-prediction of the number of objects for all three mock samples for the entire redshift range. To compute the $M_{\rm 500c}$ mass function using the MiraTitanEmulator, which predicts the $M_{\rm 200c}$ HMF, we assumed a NFW density profile and the mass-concentration relation from \citet{Diemer2019}. At the steep end of the HMF, slight deviations from the truth and scatter in the mass-concentration relation has a big effect on the results. Our results highlight the fact that it is important to use models that emulate the quantity of interest directly, as conversion factors will introduce additional biases.

The HMF model that best recovers the FLAMINGO cluster counts is the model from \citet{Bocquet2016}. From the left panel of Fig.~\ref{fig:HMF_comp} it is clear that this is due to the fact that the model agrees with the simulation at the few per cent level up to $M_{\rm 500c}\approx7\times10^{14}~\rm{M}_{\odot}$. Above this mass the model underpredicts the number of clusters. This explains the behaviour of this model in the right panel. For the Planck-like survey, the model undershoots the results, as the Planck-like survey is mostly limited to very massive clusters. For the deeper surveys, that probe objects down to a lower mass than where the model starts to deviate, there is reasonable agreement. For the SPT-like survey it still undershoots, but stays largely within the error bars. For much of the redshift range, there is good agreement with the SO-like survey, only deviating at high redshifts, though it is hard to see in this plot whether it stays within the error bars. Because of this relatively good agreement, we will assume the model by \citet{Bocquet2016} for later comparisons and fits where we need to vary the cosmology.

From this comparison of HMF models it is clear that the choice of model will have a direct influence on the inferred cosmology, and that there are big differences between the models. For $M_{\rm 500c}\lesssim2\times10^{14}~\rm{M}_{\odot}$ the models by \citet{Bocquet2016} and \citet{MiraTitanHMF2020} are in good agreement with the FLAMINGO DMO simulation. However, most objects, especially in SZ-selected surveys, will have higher masses than this. While the sensitivity of the high mass end of the HMF to cosmology makes it an interesting target, the model predictions are very sensitive to any of the choices made. This includes, for example, mass conversions, simulation box size effects, halo definitions and the halo finder used. This includes the choices we made for FLAMINGO used in this work (HBT+, $M_{\rm 500c}$). Therefore, it might be more beneficial to focus on slightly lower mass objects that suffer less from these systematic errors, something that will happen naturally with upcoming surveys.

\begin{figure*}
    \centering
        \includegraphics[width=.99\columnwidth]{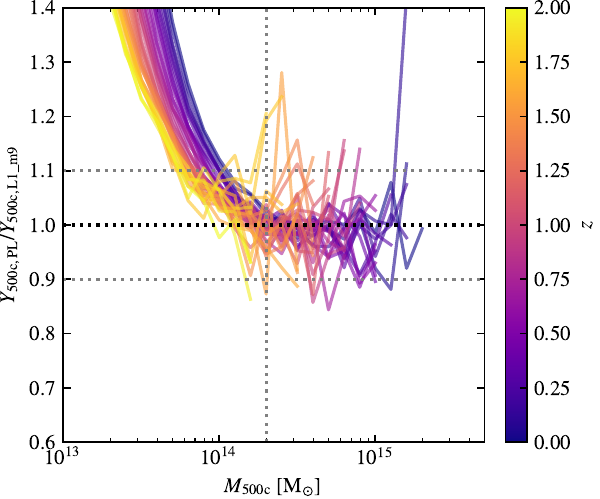}
        \includegraphics[width=.99\columnwidth]{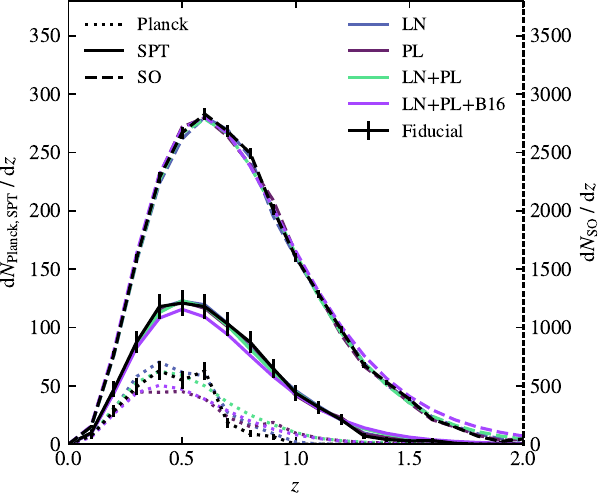}
    \caption{The left panel shows the ratio between the best-fitting power-law $Y_{\rm 500c}(M_{\rm 500c})$ to the scaling relation of the L1\_m9 simulation and the actual scaling relation at different redshifts, indicated by the different colours. The black dotted horizontal line shows the one-to-one line and the grey dotted horizontal line shows ten per cent deviations. The grey dotted vertical line indicates the lower mass limit used for fitting which roughly coincides with the lowest mass clusters that will be in the deepest (SO) survey. The right panel shows the distribution of clusters with redshift for the three different surveys, indicated by different line-styles. The different coloured lines show the effect of assuming the scaling relation to be a power-law (PL), the scatter to be lognormal (LN), the HMF to be given by the MiraTitanEmulator (MTE) and combinations of these assumptions. The black line with error bars shows the cluster counts from the fiducial model (L1\_m9 observable-mass scaling relation and L5p6\_m10\_DMO HMF). The error bars are the Poisson errors in each bin. Note that the left y-axis is for the dotted (Planck) and solid (SPT) lines, while the right y-axis is for the dashed lines (SO).}
    \label{fig:PL_comp}
\end{figure*}

\subsection{Baryonic effects on the halo mass function}\label{sec:barhmf}
In addition to biases introduced by the choice of DMO HMF model, all of the models neglect the effect of baryons on the halo masses. If the observable-mass scaling relation is obtained from hydrostatic masses \citep{PlanckClustercosmo2016} then it is the total mass, that includes both baryons and dark matter, that is probed. If the halo baryon fraction deviates from the universal fraction, then this leads to an inconsistency between the scaling relation and the HMF used for modelling. This can be solved by converting weak lensing masses to DMO halo masses by using matched haloes between hydro and DMO simulation in the same volume \citep{Grandis2021}, but this procedure still depends on accurate modelling of the baryons. Moreover, the total halo masses inferred from weak lensing depend on assumed density profiles and are therefore themselves subject to baryonic effects \citep{Debackere2021}. For the largest objects the effect of baryons on the HMF is predicted to be small, but it is expected to increase towards lower masses \citep[see e.g.][]{Velliscig2014,Cui2014,FLAMINGOmain}. To quantify the uncertainty in the baryonic effect on the HMF, we will make use of the FLAMINGO feedback variations, which span the uncertainties in the observed gas fraction data \citep{Kugel2023}.

The impact of the baryonic effects on the HMF is shown in Fig.~\ref{fig:HMF_bar}. Similar to Fig.~\ref{fig:HMF_comp}, the left panel shows the ratio between the HMF of the variations and the L1\_m9\_DMO simulations at $z=0$. We choose to compare with the L1\_m9\_DMO instead of the L5p6\_m10\_DMO as these simulations use the same initial conditions and therefore provide a fairer comparison when investigating just the effect of baryons. Baryonic effects suppress the mass function more strongly at lower masses. For the L1\_m9 model the suppression decreases from $\approx 20$ per cent at $M_{\rm 500c} \sim 10^{13}~\rm{M}_{\odot}$ to $\approx 10$ per cent at $10^{14}~\rm{M}_{\odot}$. As expected, models with lower gas fractions lead to stronger suppression. 

A $(1~\rm{Gpc})^3$ volume only contains a limited number of objects with $M_{\rm 500c}>10^{15}~\rm{M}_{\odot}$. Therefore, the ratio becomes quite noisy for the highest masses, as can be seen in the left panel of Figure~\ref{fig:HMF_bar}. As noted in Section~\ref{sec:cluster_count_model}, we set the ratio to one when either of the bins (in the DMO or the hydro simulation) is empty, which can be seen for the largest masses. As discussed in the previous section, the high-mass end is the part of the HMF where systematic effects become dominant. The large amount of noise has some effect on the Planck-like survey, but it is negligible for the other surveys.

The right panel shows how the deviations from DMO, shown in the left panel, propagate into changes in the cluster counts. Because the ratio is noisy for the highest masses, we can see in the right panel that some of the models fall outside the error bars for the Planck-like survey. The fact that we find disagreement for the sample that covers the highest mass clusters indicates that we are dominated by small number statistics due to the limited simulation volume. Therefore, we choose not to make any definite statements on whether the effect of baryons on the HMF would lead to a large bias for a Planck-like sample.

\begin{figure*}
    \centering
        \includegraphics[width=.99\textwidth]{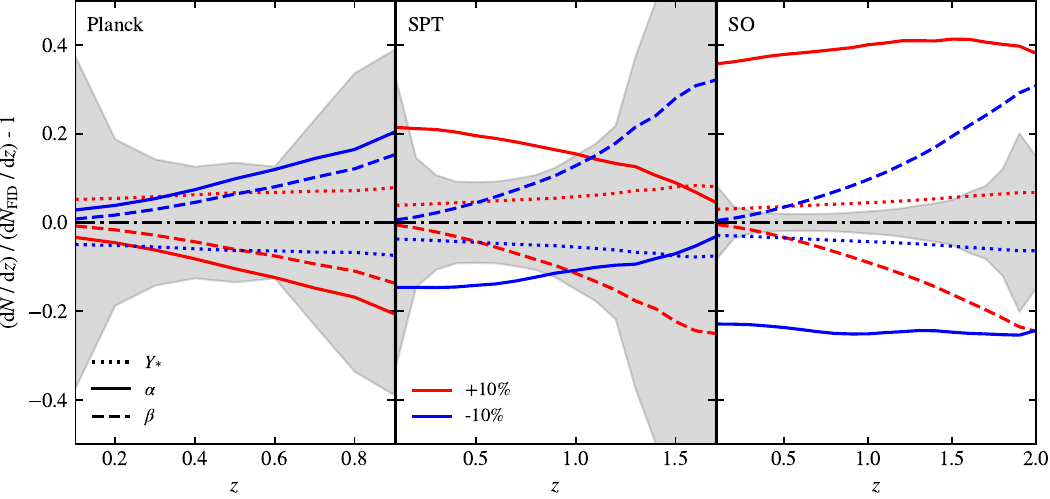}
    \caption{The relative impact of changes to the power-law parameters to the cluster counts. From left to right, the three panels show the impact for the Planck-, SPT- and SO-like surveys respectively. The different line-styles indicate the three different power-law parameters (Eq.~\ref{eq:powerlaw}): the amplitude $Y_{*}$, the scaling with mass $\alpha$ and the redshift scaling $\beta$. The red (blue) lines increase indicate an increase (decrease) by $10\%$ relative to the fiducial value of each parameter. The black dot-dashed line indicates the one-to-one line. The grey band indicates the Poisson errors on each sample. Accuracy higher than 10\% is needed for every parameters for SPT and SO. The sensitivity to the mass scaling increases for deeper surveys.}
    \label{fig:PL_uncertainties}
\end{figure*}

For the SPT-like survey most variations fall within the Poisson error bars, only the most extreme models, fgas-8$\sigma$ and Jet\_fgas-4$\sigma$, fall outside the error bars. These models start to deviate by over $10\%$ close to the mass cut. For the SO-like survey, nearly every baryonic modification leads to disagreements beyond the Poisson error bars over most of the redshift range, especially at $z\approx0.5$. Only the weakest feedback model has a small enough impact to still be consistent with the fiducial model. This can also be seen in the left panel, where the fgas+2$\sigma$ model is consistent within a few per cent for the masses probed by the SO mock sample, indicated with a vertical dotted line. We conclude that future observations cannot neglect the effect that baryons have on the HMF.

\subsection{Fit to the observable-mass scaling relation}\label{sec:models}
The next two subsections explore the effect of varying the assumptions for the mass-observable scaling relation, i.e. the relation between $M_{\rm{500c}}$ and $Y_{\rm 500c}$, and its scatter. In this section, we investigate the effect of assuming a power-law with lognormal scatter to model the scaling relation. In subsection \S\ref{sec:scalvars} we explore the effect of getting the scaling relation slightly wrong by making use of the FLAMINGO feedback variations.

The power-law parameterization we use is described in Section~\ref{sec:cluster_count_model} (Eq.~\ref{eq:powerlaw}) and is taken from \citet{PlanckClustercosmo2016}. Instead of using their best-fitting values, we re-fit the amplitude $Y_{*}$, the slope $\alpha$, the redshift scaling $\beta$ and the lognormal scatter $\sigma$. We re-fit to isolate the effects of assuming a power-law functional form. We only fit these parameters for masses $M_{\rm 500c}>2\times10^{14}~\rm{M}_{\odot}$. The result of this can be seen in the left panel of Fig.~\ref{fig:PL_comp}. The re-fit ensures that the relation leads to a match for the masses probed by our three surveys, across all redshifts, though extrapolation to lower cluster masses would result in large errors. The results of the fit are listed in Table~\ref{tab:params}. Note that we use a fit that extends to much lower masses than required for the Planck-like survey. If we would self-calibrate the relation using only objects that would be detected, the fits to the Planck- and SPT-like surveys would likely degrade.

If we compare the values we find with those given in literature, we find that the amplitude $Y_{*}$ and the slope $\alpha$ both have to change by more than 1$\sigma$. The redshift scaling $\beta$ is within the 1$\sigma$ range. However, we note that changing $\beta$ within the 1$\sigma$ range can lead to large changes in the predicted cluster counts. The value $\beta=0.89$ that we find for FLAMINGO is also significantly different from the self-similar scaling, $\beta=2/3$. The amplitude $Y_{*}$ is degenerate with the hydrostatic mass bias $b$, hence any small differences between the hydrostatic bias used and inferred by Planck, and the value we use that was found during calibration, can quickly lead to differences in the values of the amplitude $Y_{*}$. The parameters $\beta$ and $b$ are both kept fixed in the analysis by \cite{PlanckClustercosmo2016}. To get rid of the degeneracy due to the hydrostatic bias we assume a fixed value given by the FLAMINGO calibration by \cite{Kugel2023}. However, we note that using the value of the hydrostatic bias used by \cite{PlanckClustercosmo2016} ($b=0.780$) has a big impact on the resulting value of $Y_{*}$. The log-normal scatter we find for FLAMINGO, $\sigma_{\log_{10}}=0.081$, is consistent with the Planck prior, $\sigma_{\log_{10}}=0.075\pm0.01$. As the scatter for Compton-Y within $R_{\rm 500c}$ is close to lognormal \citep{Kugel2024}, any deviations found here are more likely to be an effect of the scatter not being constant with mass or redshift.

While we obtain a good match to the Compton-Y values for the masses we fit to, we find that the scaling relation quickly starts to deviate at lower masses, indicating that a double power-law might better describe the scaling of the integrated SZ signal with mass. However, with current facilities we cannot probe these masses efficiently observationally, so we do not attempt to constrain a double power-law. Above the mass cut we match the relation up to redshift two.

\begin{figure*}
    \centering
        \includegraphics[width=.99\columnwidth]{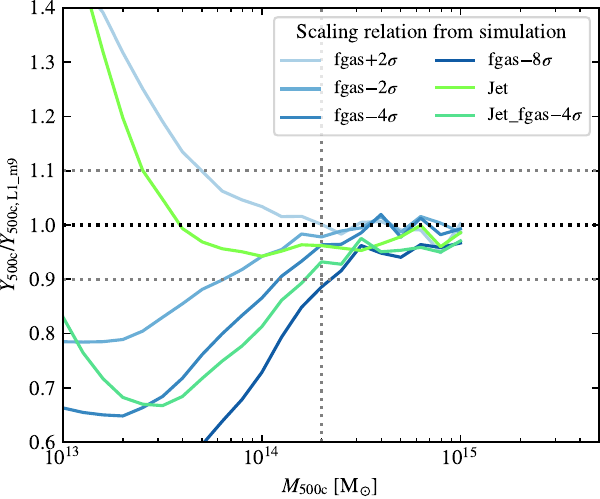}
        \hspace{0.5 cm}
        \includegraphics[width=.99\columnwidth]{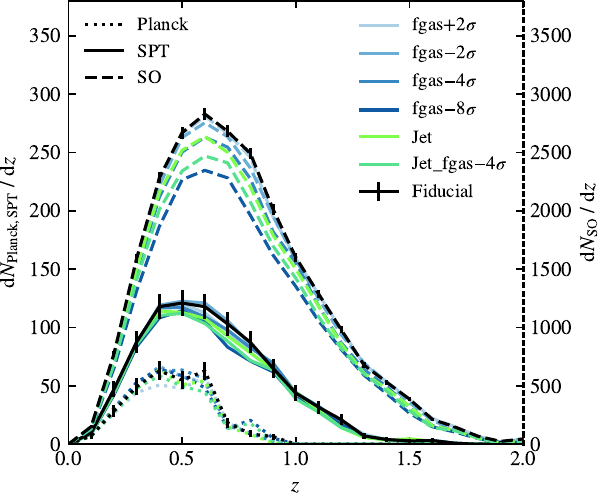}
    \caption{The left panel shows the ratio of the $Y_{\rm 500c}(M_{\rm 500c})$ scaling relation of the different FLAMINGO feedback variations, listed in the legend, and that of the fiducial L1\_m9 simulation at $z=0$. The black dotted horizontal line shows the one-to-one line and the dotted grey horizontal line show ten per cent deviations. The grey dotted vertical line roughly coincides with the lowest masses that will be in the deepest (SO) sample. The right panel shows the distribution of clusters with redshift for the three different surveys, indicated by different line-styles. The different coloured lines show the effect changing the scaling relation to one of the variations has for each FLAMINGO simulation. The black line with error bars shows the cluster counts from the fiducial model (L1\_m9 observable-mass scaling relation and L5p6\_m10\_DMO HMF). The error bars are the Poisson errors in each bin. Note that the left y-axis is for the dotted (Planck) and solid (SPT) lines, while the right y-axis is for the dashed lines (SO).}
    \label{fig:scale_vars}
\end{figure*}

The impact of the assumptions for the scaling relation on the cluster counts can be seen in the right panel of Fig.~\ref{fig:PL_comp}. The different abbreviations stand for lognormal scatter (LN), power-law (PL) and using the model by \citet{Bocquet2016} (B16) for the HMF. For the Planck-like survey, the power-law leads to a substantial decrease in the number of objects across most redshifts. At $z>0.7$, all assumptions lead to large overestimates. Just like for the HMF modification, it is likely that this behaviour is in large part due to the limited statistics for very massive clusters in the L1\_m9 simulation. For the SPT-like survey we see that the agreement is generally very good. At nearly all redshifts, the lognormal scatter and power-law assumptions have a negligible effect. Only when combined with the HMF model  from \citet{Bocquet2016} (LN+PL+B16), which we know from Fig.~\ref{fig:HMF_comp} leads to deviations, do we see a noticeable impact. As shown in a previous section, this is driven by errors in the HMF.

For the SO-like survey, the relative differences seen in the plot are very small, and there are almost no systematic offsets. However, as the sample is much larger, the requirements for a hypothetical cosmological parameter inference to be accurate down to cosmic variance levels are also much stricter. The largest deviation is caused by the introduction of the HMF model by \citet{Bocquet2016}, as expected from section~\ref{sec:hmf_comp}. This is seen for $z>1$ where there is an over-prediction of the number of clusters. Other deviations can be seen around the peak of the distributions, where all the assumptions lead to an underestimate that is outside the Poisson error bars either before the peak, for LN, or after the peak, for PL and LN+PL. As discussed in Section~\ref{sec:biasquant}, these minute differences still lead to larger errors than those due to cosmic variance.

Provided the observed scaling relations can be constrained in an unbiased way, these results show that the commonly assumed power-law functional form is unlikely to lead to a large bias, at least not for surveys that are only sensitive to masses $M_{\rm 500c} > 2\times10^{14}~\rm{\odot}$. However, the values of the parameters of the scaling relation will need to be fit with high accuracy. To investigate what accuracy is needed, we look at the effect of changing each parameter in Fig.~\ref{fig:PL_uncertainties}. In contrast with previous figures, we show the relative change with respect to a model that uses the fiducial values for the fitting parameters. In other words, the difference between the one-to-one line and the coloured lines is only due to the change in a single parameter. The different line-styles indicate the three different parameters we vary. The colour indicates whether the parameter is increased (red) or decreased (blue) by 10\%. The grey coloured regions indicate the Poisson errors on each survey. For reference, for the constraints used by \citet{PlanckClustercosmo2016} the relative uncertainty is about 10\% for $Y_{*}$, 5\% for $\alpha$ and 75\% for $\beta$. This is improved upon by \citet{Bocquet2024}. While they use a different definition for the scaling relations, their respective parameters for $Y_{*}$, $\alpha$ and $\beta$ have an accuracy of 8.7\%, 2.3\% and 14.8\% respectively in their final analysis.

Looking at the different line-styles in Fig.~\ref{fig:PL_uncertainties}, we can see that for a large part of the redshift range, the parameters $\alpha$ and $\beta$ have the largest effect at a fixed relative uncertainty of 10 per cent. The uncertainty in $\beta$ leads to larger errors with increasing $z$. The uncertainty in $\alpha$ leads to bigger errors as the survey gets deeper. This is as is to be expected. The functional form for the power-law has a pivot mass at $M_{\rm 500c}=6\times10^{14}~\rm{M}_{\odot}$. An error in $\alpha$ leads to an increasing bias in the scaling relation when the masses differ more from the pivot mass. The amplitude $Y_{*}$ leads to a similar level of uncertainty of about 5 to 10 per cent for each survey, slightly decreasing for the deeper surveys.

Going panel-by-panel, it is clear that a $10\%$ uncertainty in the fitting parameter only leads to an uncertainty in the cluster counts that is within the Poisson error bars for the Planck-like survey. For the SPT and SO mocks, a higher accuracy is needed for each of the three parameters. The constraints by \cite{PlanckClustercosmo2016} on the parameter $\beta$ have a much larger uncertainty (75\%) than the 10\% shown in this figure. Furthermore, a $10\%$ deviation from the best fitting value found for FLAMINGO is still inconsistent with self-similar evolution. This implies that assuming the evolution is self-similar might lead to significant biases. As the deeper surveys also probe higher $z$, it is important that the redshift evolution is captured correctly. Additionally, for deeper surveys the pivot mass should be moved to lower masses to reduce the error due to the uncertainty in $\alpha$. At the current pivot, an error in $\alpha$ of less than a per cent is required to keep the results within the SO Poisson error bars. For the SPT mock, we see that the constraints used by \citet{Bocquet2024} are not tight enough to push the uncertainties below the level of the Poisson errors. It is important to note that these parameters and their uncertainties are marginalised over when inferring cosmology and that the uncertainties on these parameters will factor into the uncertainty in the cosmology. Our results highlight the importance of fully understanding the selection effects when constraining the scaling relations, as small errors can lead to large biases in the inferred cosmology.

\begin{table*}
    \centering
    \caption{Compilation of the level of agreement between the cluster number counts predicted using the various assumptions tested in this work and those predicted by the fiducial model, which uses the L1\_m9 observable mass scaling relation and the L5p6\_m10\_DMO HMF. Note that in each case the cosmology is fixed to the true one. The first column indicates the four different types of assumptions tested, the second column specifies which model was used for that type of assumption. The final six columns give the $\chi^2$ and the $p$-value for the hypothesis that the model is consistent with the mock sample, split by the three mock samples. Every time an assumption leads to a discrepancy that has a $p$-value lower than 0.2, it is shown in boldface. Entries with a higher value lead to systematics that are consistent within the Poisson errors.}
\begin{tabular}{|l|l|r|l|r|l|r|l|}
\hline
Assumption & Model & Planck \(\chi^2\) & Planck \(p\) & SPT \(\chi^2\) & SPT \(p\) & SO \(\chi^2\) & SO \(p\) \\
\hline

DMO HMF & Tinker (2010) & 50.7 & $\boldsymbol{<10^{-6}}$ & 107.1 & $\boldsymbol{<10^{-6}}$ & 1790.7 & $\boldsymbol{<10^{-6}}$ \\
 & Bocquet (2016) & 10.7 & $\boldsymbol{1.54\times10^{-1}}$ & 2.9 & $1.00\times10^{0}$ & 68.9 & $\boldsymbol{<10^{-6}}$ \\
 & MiraTitanEmulator & 36.0 & $\boldsymbol{7.38\times10^{-6}}$ & 64.4 & $\boldsymbol{<10^{-6}}$ & 712.5 & $\boldsymbol{<10^{-6}}$ \\ \hline

Baryon effect on the HMF & L1\_m9 & 4.2 & $7.56\times10^{-1}$ & 2.1 & $1.00\times10^{0}$ & 22.4 & $2.14\times10^{-1}$ \\
 & fgas\(+2\sigma\) & 11.5 & $\boldsymbol{1.18\times10^{-1}}$ & 3.3 & $1.00\times10^{0}$ & 5.9 & $9.97\times10^{-1}$ \\
 & fgas\(-2\sigma\) & 1.7 & $9.75\times10^{-1}$ & 2.4 & $1.00\times10^{0}$ & 96.7 & $\boldsymbol{<10^{-6}}$ \\
 & fgas\(-4\sigma\) & 1.1 & $9.93\times10^{-1}$ & 3.4 & $1.00\times10^{0}$ & 226.9 & $\boldsymbol{<10^{-6}}$ \\
 & fgas\(-8\sigma\) & 0.4 & $1.00\times10^{0}$ & 9.2 & $8.68\times10^{-1}$ & 563.1 & $\boldsymbol{<10^{-6}}$ \\
 & Jet & 1.3 & $9.88\times10^{-1}$ & 2.6 & $1.00\times10^{0}$ & 80.0 & $\boldsymbol{<10^{-6}}$ \\
 & Jet\_fgas\(-4\sigma\) & 0.4 & $1.00\times10^{0}$ & 6.2 & $9.77\times10^{-1}$ & 404.0 & $\boldsymbol{<10^{-6}}$ \\ \hline

Scaling relation fit & LN & 14.5 & $\boldsymbol{4.29\times10^{-2}}$ & 2.5 & $1.00\times10^{0}$ & 35.6 & $\boldsymbol{8.03\times10^{-3}}$ \\
 & PL & 22.3 & $\boldsymbol{2.22\times10^{-3}}$ & 1.9 & $1.00\times10^{0}$ & 20.5 & $3.04\times10^{-1}$ \\
 & LN+PL & 49.1 & $\boldsymbol{<10^{-6}}$ & 13.5 & $5.66\times10^{-1}$ & 127.6 & $\boldsymbol{<10^{-6}}$ \\
 & LN+PL+B16 & 33.2 & $\boldsymbol{2.44\times10^{-5}}$ & 23.2 & $\boldsymbol{7.97\times10^{-02}}$ & 412.9 & $\boldsymbol{<10^{-6}}$ \\ \hline

Scaling relation variations & fgas\(+2\sigma\) & 4.3 & $7.48\times10^{-1}$ & 2.6 & $1.00\times10^{0}$ & 4.0 & $1.00\times10^{0}$ \\
 & fgas\(-2\sigma\) & 2.6 & $9.18\times10^{-1}$ & 0.9 & $1.00\times10^{0}$ & 31.3 & $\boldsymbol{2.69\times10^{-2}}$ \\
 & fgas\(-4\sigma\) & 2.9 & $8.90\times10^{-1}$ & 2.1 & $1.00\times10^{0}$ & 156.3 & $\boldsymbol{<10^{-6}}$ \\
 & fgas\(-8\sigma\) & 20.8 & $\boldsymbol{4.00\times10^{-3}}$ & 11.7 & $6.99\times10^{-1}$ & 762.6 & $\boldsymbol{<10^{-6}}$ \\
 & Jet & 2.3 & $9.41\times10^{-1}$ & 7.8 & $9.32\times10^{-1}$ & 191.4 & $\boldsymbol{<10^{-6}}$ \\
 & Jet\_fgas\(-4\sigma\) & 16.9 & $\boldsymbol{1.81\times10^{-2}}$ & 12.3 & $6.59\times10^{-1}$ & 462.3 & $\boldsymbol{<10^{-6}}$ \\ \hline

\end{tabular}

\label{tab:ass_sum}
\end{table*}

\begin{figure*}
    \centering
        \includegraphics[width=.99\textwidth]{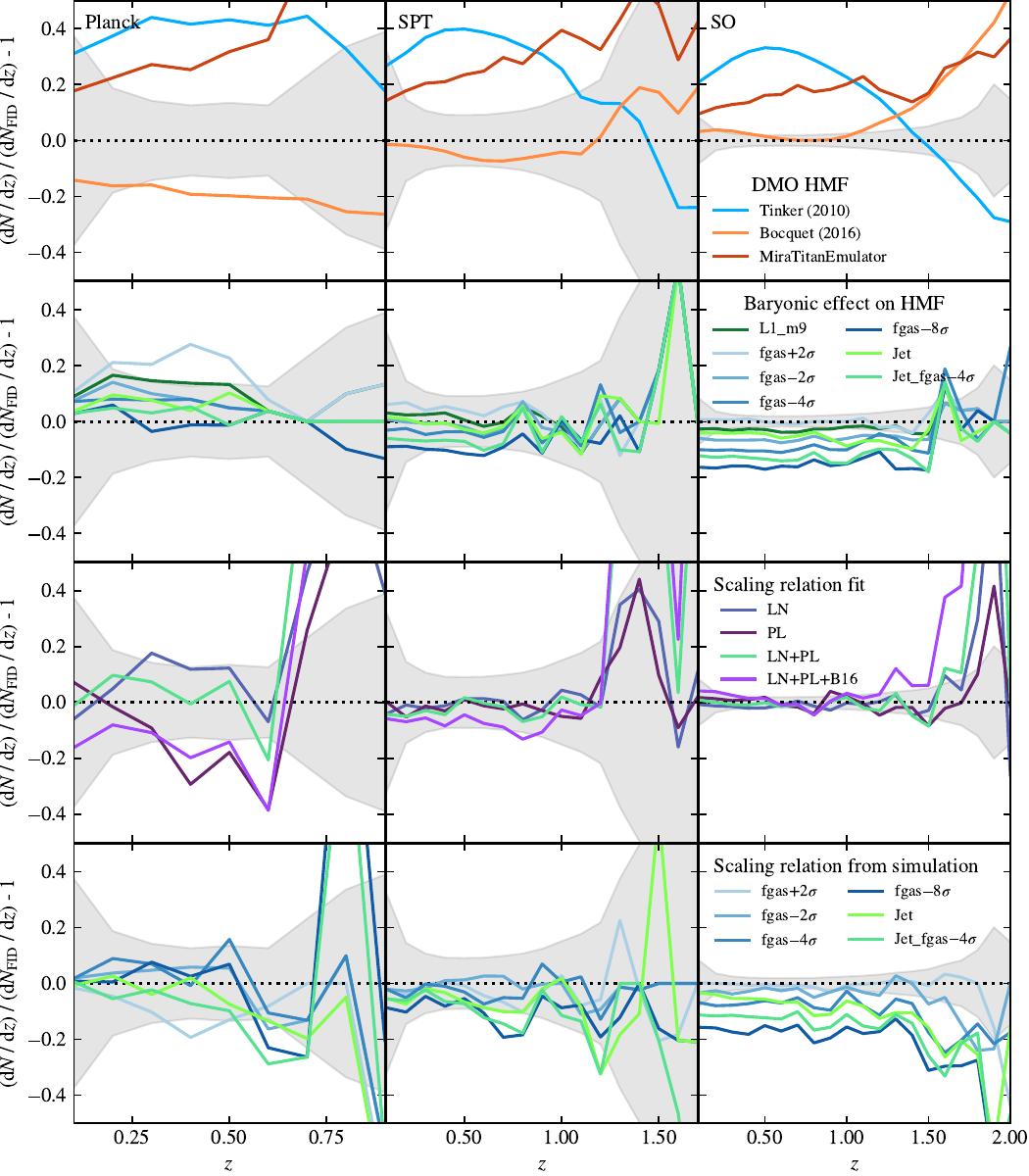}
    \caption{A compilation of all the cluster count comparisons in this work. Each panel shows the ratio between the cluster number counts predicted by a model making the indicated assumption and those predicted by and our fiducial model, which uses the L1\_m9 mass-observable scaling relation and scatter, and the L5p6\_m10\_DMO HMF. The different model assumptions are indicated by the different coloured lines. The Poisson uncertainty is given by the grey shaded region. The three columns show the results for the three mock samples considered in this work, from left to right, Planck, SPT and SO. The first row varies the model for the HMF, the second row varies the baryonic modification of the DMO HMF, the third row investigates different fits to the observable-mas relation predicted by the fiducial model and the fourth row varies the simulation from which the mass-observable scaling relation and scatter is taken. In the third row the abbreviations are lognormal (LN), power-law (PL) and using the HMF model by \citet{Bocquet2016} (B16). The biggest systematic errors for all surveys are caused by the HMF models, however, for the SO-like survey, nearly all assumptions lead to a significant errors.}
    \label{fig:big_rat_fig}
\end{figure*}

\subsection{Systematic uncertainty in the observable-mass scaling relation}\label{sec:scalvars}
In the previous section we have investigated the impact of assuming a power-law with lognormal scatter on the cluster counts and quantified how tightly the parameters of the power-law need to be constrained to get unbiased results. In this section we investigate the effect of changes to the scaling relation itself. To do this, we make use of the FLAMINGO variations. Similar to previous figures, we show both the change in the varied relation and the resulting effect on the cluster counts in Fig.~\ref{fig:scale_vars}.

In the left panel of Fig.~\ref{fig:scale_vars}, we show the ratio between the $Y_{\rm 500c}-M_{\rm 500c}$ scaling relation of each of the simulation variations, and the scaling relation from the L1\_m9 simulation used for our fiducial model at $z=0$. The right panels shows the impact of these deviations from the fiducial model on the cluster counts for our three mock samples, indicated by the different line-styles. The black lines show the fiducial model, including Poisson errors.

At $z=0$, and for masses included in all samples (the mass limit is indicated by the vertical grey line in the left panel), the changes in the scaling relations mostly fall within the $10\%$ region, but the deviations become much larger at lower masses. All models are similar for the largest objects. This is as expected, since AGN feedback is unable to offset the inflow of gas in the most massive haloes, even for the more extreme FLAMINGO variations. Towards lower masses the different variations diverge, with models with higher gas fractions having a higher integrated Compton-Y at fixed mass. There is an offset between the thermal and 'Jet' models at fixed gas fractions, indicating that the feedback model changes the resulting scaling relation's dependence on gas fraction.

Looking at the right panel of Fig.~\ref{fig:scale_vars}, the quantitative difference due to changing the scaling relation is similar to what was found for the baryon effect on the HMF. This can be partially explained by the fact that not only Compton-Y, but also $M_{\rm 500c}$ changes for each halo between the different variations. Therefore, some care needs to be taken when interpreting these results in the context of uncertainties in the gas fraction. For Planck most models are quite close to the Poisson error bars, and at high $z$ they are likely affected by noise as they do not deviate systematically. For SPT, A large fraction of the models falls within the Poisson error bars. Only the models with the lowest gas fractions deviate significantly, especially for $z\approx0.75$. It is only for the SO-like mock sample that almost every variation leads to a significant difference. Only the fgas+2$\sigma$ model agrees with the fiducial model within the Poisson error bars.

Assuming that deviations from the models used for cluster cosmology are as small as for our mildest feedback variations, it seems that for previous and current generation surveys the difference is still small enough for the effect to be within the Poisson error bars. However, for future surveys, the necessity of constraining the scaling relation more tightly is clear.
\begin{figure*}
    \centering
        \includegraphics[width=.99\textwidth]{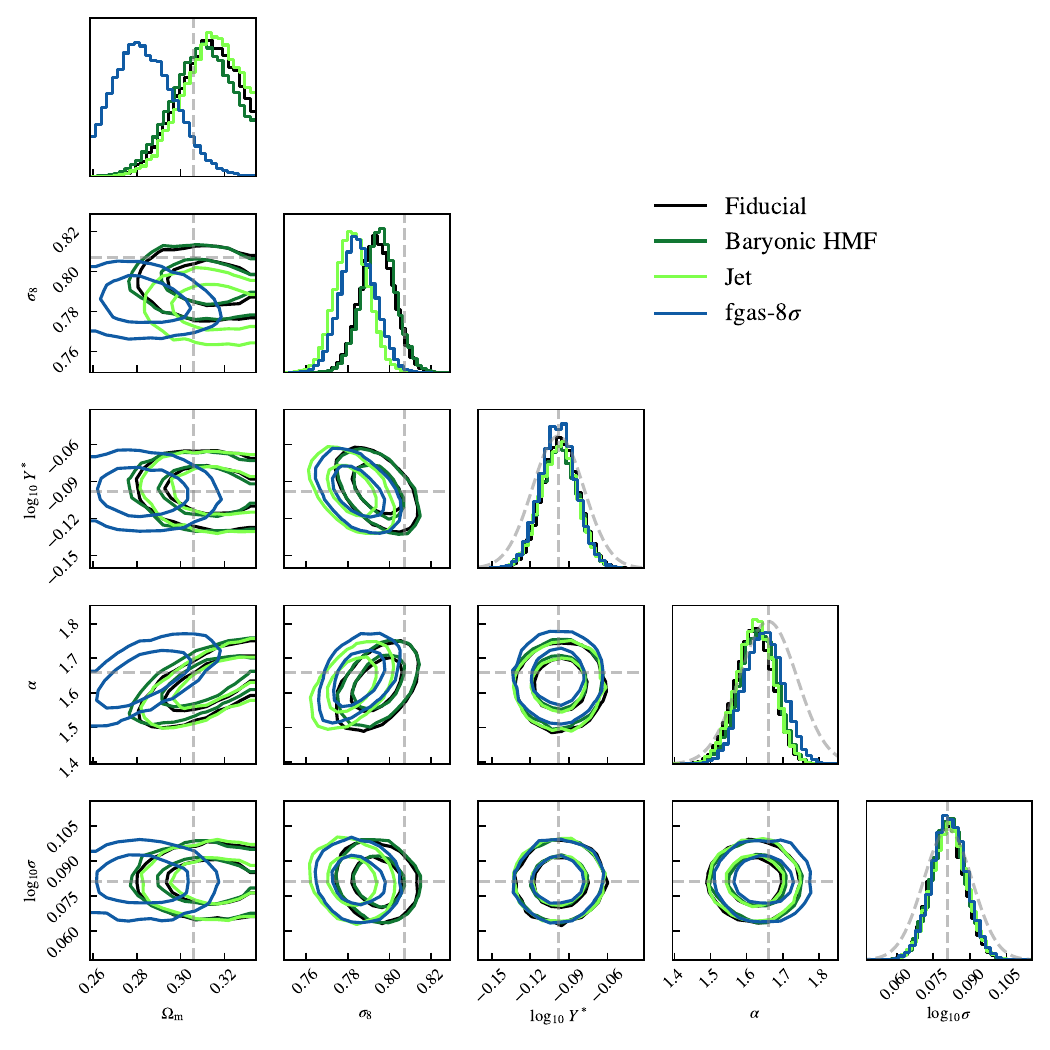}
    \caption{Posteriors from fitting the "Full" fitting model for cluster counts, which assumes a power-law mass-observable scaling relation with lognormal scatter and the DMO HMF model from \citet{Bocquet2016}, to the cluster count predictions for four different mocks setups at the Simons observatory detection limit: Our fiducial mock sample ("Fiducial"; L1\_m9 observable-mass scaling relation and scatter and a L5p6\_m10\_DMO HMF), a mock sample that includes the baryonic modification of the HMF ("Baryonic HMF"; as the Fiducial model but with the baryonic modification of the HMF predicted by L1\_m9), a setup that uses an alternative model for AGN feedback that employs kinetic jets instead of thermally-driven winds ("Jet"; the mass-observable scaling relation, scatter and baryonic modification of the HMF are taken from the Jet simulation) and a setup that uses the simulation with the strongest reduction in cluster gas fractions ("fgas-8$\sigma$"; the mass-observable scaling relation, scatter and baryonic modification of the HMF are taken from the fgas-8$\sigma$ simulation). We fit for the cosmological parameters $\Omega_{\rm m}$ and $\sigma_8$, the two power-law observable-mass scaling relation parameters $Y^*$ and $\alpha$ (see Eq.~\ref{eq:powerlaw}), and the lognormal scatter around this relation $\sigma_{\log_{10}}$. The horizontal and vertical grey dashed lines show the true values. In the diagonal panels for the parameters $Y^*$, $\alpha$ and $\sigma_{\log_{10}}$, we also show the priors using grey dashed lines. The two contour levels show the 68th and 95th percentiles. There are significant biases in the cosmological parameters for all mocks.}
    \label{fig:cornerplot}
\end{figure*}

\subsection{Quantifying the biases in the number counts}\label{sec:biasquant}

In this section we will investigate and compare the effects on the cluster number counts of the different assumptions that we investigated for the three different mock samples. This way we can get a good overview for all the assumptions that might lead to biases in the inferred cosmology.

In Table~\ref{tab:ass_sum} we compile the effect of each assumption in terms of $\chi^2$ and the $p$-value with respect to our fiducial model, which consists of the observable-mass scaling relation from the L1\_m9 simulation with the L5p6\_m10\_DMO HMF. To highlight the problematic cases, all deviations that have a $p$-value below 0.2 are highlighted using boldface. Like in the previous sections, the different assumptions are divided into four categories: the DMO HMF model, baryonic effects on the HMF, assumptions made when fitting the observable-mass scaling relation and variations in the scaling relation itself. Additionally, we summarise all the findings so far in Fig.~\ref{fig:big_rat_fig}. This figure combines the right panels of Figs~\ref{fig:HMF_comp}, \ref{fig:HMF_bar}, \ref{fig:PL_comp} and \ref{fig:scale_vars}, but instead shows the ratio of the cluster number counts with respect to our fiducial model. The three mock samples are shown in the three columns with their Poisson errors shown as the grey shaded region.

By focusing on the boldfaced entries in Table~\ref{tab:ass_sum}, we get a clear overview of the findings that we reported earlier in this work. There are a few groups of problematic assumptions to highlight. One of these groups is found for some of the results for the Planck-like survey. As explained earlier, these deviations are mostly due to the limited $(1~\rm{Gpc})^3$ volume used for this work, which leads to additional noise for the cluster masses found by the Planck-like survey.

The second group consists of the DMO HMF models, which all lead to significant biases for all surveys, with the one exception being the model by \citet{Bocquet2016} for the SPT-like survey. This can also be seen in Fig.~\ref{fig:big_rat_fig}. In all other cases the models fall outside the error bars for most of the redshift range. For the SO-like survey the requirements are particularly stringent. The model by \citet{Bocquet2016} agrees very well for a large part of the redshift range, but outside of that range the deviations quickly become so large that it leads to a $p$-value that indicates a high level of disagreement.

Finally, nearly the entire SO column is boldfaced. None of the DMO HMF models reproduce our fiducial model sufficiently well. All FLAMINGO variations but the ones with the weakest feedback have enough of an effect on the HMF and scaling relation to lead to significant differences. The assumption of lognormal scatter on the mass-observable scaling relation, especially in combination with the assumption that the scaling relation is a power-law and the previously mentioned inaccuracies of the DMO HMF model assumptions, lead to large deviations. From Fig.~\ref{fig:big_rat_fig} it is clear that for the baryonic variations, the effects on the HMF and on the observable–mass scaling relation are insensitive to redshift, leading to systematic offsets across the entire redshift range. For the lognormal scatter and power-law assumptions, the discrepancies are greatest at $z>1$. These findings highlight the point that for future surveys our current models use assumptions that are too restrictive.

Our findings are in line with those of \citet{Bocquet2016}, who also find that post-SPT like surveys need to take baryonic corrections into account, and that we need to be careful in constructing the HMF models as different current models in the literature will lead to the inference of significantly different cosmologies. 

\subsection{Biases in cosmological parameters}\label{sec:fitall}
As a final test we want to quantify the impact of the assumptions that are commonly made in models predicting cluster counts on inferred cosmological parameters for SPT- and SO-like surveys. We exclude the Planck-like survey from this section as we found previously that the systematics are limited by our finite simulation volume. 

To fit mock sample data, we make use of the "full" model, LN+PL+B16, which assumes the DMO HMF model from \citet{Bocquet2016}, the power-law observable-mass scaling relation given by Eq.~\ref{eq:powerlaw} with lognormal scatter. We fit for five parameters: the cosmological parameters $\Omega_{\rm m}$ and $\sigma_{8}$, the power-law observable-mass scaling relation parameters $Y_{*}$ and $\alpha$, and the lognormal scatter in the scaling relation $\sigma_{\log_{10}}$. For the cosmological parameters we use the flat priors given by the limits of the MiraTitanEmulator $[\Omega_{\rm m} = \mathcal{U}(0.259,0.334),\sigma_{8}=\mathcal{U}(0.7,0.9)]$. For the power-law and scatter parameters we use Gaussian priors taken from \citet{PlanckClustercosmo2016} centered on the values that best fit our fiducial simulation L1\_m9 $[\log_{10}{Y_{*}}=\mathcal{N}(-0.098,0.02),\alpha=\mathcal{N}(1.66,0.08),\sigma_{\log_{10}}=\mathcal{N}(0.081,0.01)]$. We keep the redshift scaling $\beta$ and the hydrostatic bias $b$ fixed [$\beta=89,b=0.0743$]. See also Table~\ref{tab:params}. 

Besides our full fitting model, we use another fitting model, referred to as "HMF", that also assumes the DMO HMF from \citet{Bocquet2016}, but uses the actual L1\_m9 observable-mass scaling relation and scatter, rather than the power-law fit with lognormal scatter used for the full model. This model will aid in separating the errors induced by the use of the DMO HMF model from \citet{Bocquet2016} from the errors induced by the assumption of a power-law scaling relation with lognormal scatter. In the HMF case we only fit for the cosmological parameters $\Omega_{\rm m}$ and $\sigma_8$.

For each type of survey (SPT- or SO-like), we use four different mock sample setups to quantify the effect of biases in the fitting models. The first setup uses the L1\_m9 scaling relation and scatter (rather than a power-law fit and lognormal scatter) and the L5p6\_m10\_DMO HMF. This mock is fit using both the HMF and full models. As this mock setup and the fitting models both use a DMO HMF, these two fits help us quantify errors due to the choice of DMO HMF model, and errors due to the assumption of a power-law observable-mass scaling relation with lognormal scatter. 

The second mock setup, referred to as "baryonic HMF", again uses the L1\_m9 scaling relation and scatter and the L5p6\_m10\_DMO HMF, but we modify the DMO HMF by the baryonic effect predicted by the L1\_m9 simulation, which we fit using the full fitting model (LN+PL+B16). This way we can quantify the level of bias introduced by baryonic effects on the HMF not being taken into account by the fitting model. 

The final two mock setups, which we fit using the full fitting model, use the L5p6\_m10\_DMO HMF, but the observable-mass scaling relation, scatter and baryonic modification of the HMF are taken from either the Jet or fgas-8$\sigma$ simulations. These last two mocks can inform us about the impact of inconsistencies between the priors in the observable-mass scaling relation, which are centered on the predictions from L1\_m9, and reality (assumed to be either the Jet or fgas-8$\sigma$ predictions). In the case of the Jet model these inconsistencies are due to a change in the prescription for AGN feedback at fixed cluster gas fractions, while for fgas-8$\sigma$ they are due to a large reduction in the gas fractions. This yields ten fits in total, five each for the SPT- and SO-like surveys.

The results are shown in Table~\ref{tab:some_vals}. The full model can only reach the level of accuracy necessary to improve on the \citet{Planck2020} errors (i.e.\ about $2\%$ in $\Omega_{\rm m}$ and $1\%$ in $\sigma_8$) for the Fiducial mock setup for the SPT-like survey. The agreement is worst for the Jet and fgas$-8\sigma$ mock setups. For the SO mocks the fractional systematic errors are generally reduced with respect to the SPT mocks, but, because of the increased precision reached by the SO-like survey, the biases are more more significant, i.e.\ the errors correspond to a larger number of $\sigma$. 

Interestingly, for the SO survey the biases are less significant for the baryonic HMF mock setup than for the Fiducial setup. As the full fitting model does not take baryonic effects on the DMO HMF into account, this implies that unaccounted for systematic errors are compensating each other. For all fits using the full model the number of $\sigma$ is close to or exceeds unity, signaling that all models tested suffer from significant systematics. Comparing the biases for the full and HMF fitting models, it is clear that the DMO HMF model drives part of the errors for both surveys. For the SPT survey about half of the fractional error in $\Omega_{\rm m}$ and a quarter of the fractional error in $\sigma_{8}$ are caused by the DMO HMF model. For SO it accounts for most of the $\Omega_{\rm m}$ error and about half of the $\sigma_8$ error.

The posteriors for the SO-like model fits are shown in Fig.~\ref{fig:cornerplot}. Focusing first on the power-law and scatter parameters, it is clear that they are all mostly prior driven. When using broad flat priors, the posteriors for the power-law and scatter parameters posteriors extend far beyond realistic ranges. The posteriors are somewhat tighter than the priors, indicated by the dashed grey lines in the diagonal panels, but the differences are small. There are some degeneracies between the power-law and cosmological parameters. Such degeneracies can lead to biases. This can be seen for $\alpha$, which is biased slightly low due to its degeneracy with $\sigma_8$. This illustrates the importance of priors, i.e.\ independent constraints on the scaling relation parameters, in order to reduce the biases that can be incurred due to movement along the lines of degeneracy. It is also clear from the fits to the Jet and fgas-$8\sigma$ mocks that changes in the observable-mass scaling relation are not picked up by the fitting. The prior is so dominant that any differences between the prior scaling relation and the truth will manifest itself in a change in cosmological parameters. Relaxing the priors might aid in relative consistency, but this would also degrade the cosmological constraining power.

Clearly, improvements are needed to the modelling of both baryonic effects and the DMO HMF. With current models, the fractional systematic errors exceed the current Planck constraints. Without improvements on the modelling side, we will not be able to use cluster counts to make any definite statements about for example the $\sigma_8$ tension.
\begin{table}
    \centering
    \caption{The fractional and number of $\sigma$ (relative to Poisson) systematic error (see Section~\ref{sec:likeli}) in the cosmological parameters $\Omega_{\rm m}$ and $\sigma_8$ when fitting different models to the cluster counts predicted by the different mock setups for SPT- and SO-like mock samples. The first column lists the survey from which the selection limit and sky area are taken, either SPT or SO. The second column lists the model that is fit to the mock data. "HMF" uses the true observable-mass scaling relation and scatter from the fiducial simulation (L1\_m9) and the DMO HMF from \citet{Bocquet2016}. For this fitting model the only free parameters are $\Omega_{\rm m}$ and $\sigma_8$. "Full" also uses the B16 DMO HMF, but assumes a power-law (2 free parameters) plus lognormal scatter (1 free parameter) scaling relation. The third column lists the setup of the mock samples that the models are fit to. "Fiducial" stands for the true L1\_m9 scaling relation and scatter combined with the L5p6\_m10\_DMO HMF. "Baryonic HMF" uses the Fiducial setup but modifies the DMO HMF with the baryon response predicted by L1\_m9. The Jet and fgas-8$\sigma$ setups use the scaling relation, scatter and baryonic modification of the HMF from their respective simulation, with the L5p6\_m10\_DMO HMF. The fourth and fifth columns list the per cent error in the median of the cosmological parameters obtained from the fits. The final column shows the number of sigma the $\Omega_{\rm m}$-$\sigma_8$ posterior medians are away from the truth, taking into account degeneracies between the parameters.}
    \begin{tabular}{l|l|l|r|r|r}
    \hline Survey & Fit model & Mock setup & $\Omega_{\rm m}~[\%]$ & $\sigma_{8}~[\%]$ & $N\sigma$ \\ \hline
      SPT & HMF & Fiducial & 1.8 & 0.7 & 0.40\\
      SPT & Full & Fiducial & 3.8 & 3.2 & 1.38\\
      SPT & Full & Baryonic HMF & 7.6 & 5.6 & 1.36\\
      SPT & Full & Jet & 9.3 & 1.7 & 0.96\\
      SPT & Full & fgas-8$\sigma$ & 17.3 & 10.6 & 1.12 \\ \hline 
      SO & HMF & Fiducial & 2.3 & 0.7 & 1.51 \\
      SO & Full & Fiducial & 2.1 & 1.6 & 1.74 \\
      SO & Full & Baryonic HMF & 1.3 & 1.5 & 1.43\\
      SO & Full & Jet & 2.6 & 3.1 & 3.17\\
      SO & Full & fgas-8$\sigma$ & 7.5 & 2.8 & 3.54 \\ \hline
    \end{tabular}
    \label{tab:some_vals}
\end{table}

\section{Conclusions}\label{sec:conc}

Galaxy cluster counts have the potential to provide an alternative avenue to explore cosmological tensions between the cosmological parameters measured by different types of observables. For cluster surveys, selection based on the amplitude of the thermal Sunyaev-Zel'dovich (SZ) effect, i.e.\ the Compton-Y parameter, is predicted to result in smaller biases relatively to a mass-selected sample compared with e.g.\ X-ray selection \citep[e.g.][]{Kugel2024}. The number of objects detected in SZ is rapidly increasing \citep{PlanckClustercosmo2016,Hilton2018,SPTclustercosmo2019,Bocquet2024} and future observatories are predicted to dramatically increase our sample size \citep[e.g.][]{SimonsObs2019}. To exploit the statistical power of these surveys, any biases in the predictions of the models must be small compared with the statistical errors. 

In the light of these advances we investigated some of the assumptions that go into the models used to measure cosmological parameters using galaxy cluster samples, particularly those from SZ surveys. These models need to assume a model for the halo mass function (HMF), which is typically based on a framework built on dark matter only (DMO) simulations, which is conventionally combined with a power-law mass-observable scaling relation with lognormal scatter. We aim to investigate whether these model ingredients, in particular the choice of DMO HMF model, the assumptions of a power-law scaling relation with lognormal scatter, and the uncertainties in baryonic effects on the HMF and on the scaling relation, lead to a biased cosmological inference.

We make use of the FLAMINGO suite of hydrodynamical simulations \citep{FLAMINGOmain,Kugel2023}. With its large box sizes (at least 1 Gpc on a side) and many variations spanning the uncertainties in the strength of feedback processes constrained by observations of the galaxy mass function and cluster gas fractions, it provides an ideal laboratory to compare cluster count models with. The FLAMINGO simulations self-consistently predict the observables needed for cluster counts, in our case the Compton-Y parameter. We take the predictions of a FLAMINGO simulation as the baseline truth, and test the validity and implications of the different assumptions that go into the theoretical models for cluster counts that are used to measure cosmological parameters from observed cluster counts. We construct three mock samples based on the number of objects in the Planck \citep{PlanckClustercosmo2016}, SPT \citep{Bocquet2024} and future Simons observatory \citep{SimonsObs2019} SZ surveys. Since we are focused on uncertainties on the modelling side, we to not create virtual observations and assume pure Poisson errors based on the depth and area of the survey. To identify potential biases, we compare the systematic deviations to the Poisson errors of each survey. We summarise our main results in Fig.~\ref{fig:big_rat_fig} and Table~\ref{tab:ass_sum}. Our main results are:
\begin{itemize}
    \item There are large and highly significant deviations between the predictions of widely used DMO HMF models. The models differ both from each other and from the $(5.6~\rm{Gpc})^3$ DMO FLAMINGO simulation (see Fig.~\ref{fig:HMF_comp}). For all three mock samples, the results from \citet{Tinker2010} and the MiraTitanEmulator \citep{Bocquet2024} fall outside the Poisson error bars. The model from \citet{Bocquet2016} performs least badly, but it still diverges from FLAMINGO at the high mass end, leading to disagreement for the Planck-like sample, and it is not accurate enough for the SO-like sample.
    \item The effect of baryons on the HMF will need to be accounted for. This is particularly important for an SO-like survey (see Fig.~\ref{fig:HMF_bar}), for which even some of our weaker feedback variations lead to deviations that exceed the Poisson errors. 
    \item The assumption of a power-law with lognormal scatter does not lead to large systematic errors, at least not for surveys that are only sensitive to masses $M_{\rm 500c} > 2\times10^{14}~\rm{\odot}$ (see Fig.~\ref{fig:PL_comp}). For deeper surveys a single power-law is likely inadequate. For the Planck- and SPT-like samples, the systematic errors due to these assumptions are smaller than Poisson. For the SO-like sample, the assumption of a constant lognormal scatter leads to significant deviations, which become worse when combined with the power-law assumption.
    \item Ten per cent uncertainties on the power-law parameters of the observable-mass scaling relation are only good enough for the Planck-like survey (see Fig.~\ref{fig:PL_uncertainties}). For SPT and particularly for SO, (much) tighter constraints are required to push the systematic deviations below the Poisson errors. For these surveys current constraints on the scaling relation are insufficiently tight to reduce the systematic deviation below the Poisson errors.
    \item The effect of changing the scaling relation is similar in magnitude to that of the baryonic modification of the HMF (see Fig.~\ref{fig:scale_vars}). For Planck- and SPT-like surveys, only the most extreme models fall outside the Poisson errors. However, for the SO-like sample the differences are large with respect to the Poisson errors. As the changes in the scaling relation due to residual uncertainties in the observationally-constrained galaxy formation physics are at the 10 per cent level, this further reinforces the fact that improving constraints on the observable-mass scaling relation is necessary to ensure the cosmology inference remains unbiased.
    \item Comparing all the results (see Fig.~\ref{fig:big_rat_fig} and Table~\ref{tab:ass_sum}), we find that the largest source of systematic error on the number counts is the DMO HMF modelling, which leads to a significant bias for each survey. Additionally, for future surveys, improvements will be needed across all the modelling ingredients, for example by introducing a mass- and redshift-dependent scatter and by obtaining better external constraints on the mass-observable relations and the effects of baryons on the HMF.
    \item When fitting cluster counts with a model assuming a power-law observable-mass scaling relation with lognormal scatter, with Gaussian priors taken from \citet{PlanckClustercosmo2016} but centered on the values that best fit our fiducial simulation L1\_m9, and the DMO HMF from \citet{Bocquet2016}, we find significant errors in the cosmological parameters for all mock samples (see Fig.~\ref{fig:cornerplot} and Table~\ref{tab:some_vals}). Because cluster counts on their own do not constrain the power-law parameters, any systematic deviations in the prior on the scaling relation lead directly to biases in the cosmological parameters. At the current level of model accuracy, the systematic errors are so large that cluster counts cannot be used to shed light on the $\sigma_8$ tension.
\end{itemize}

We have used the FLAMINGO simulations to highlight some of the shortcomings of current cluster count cosmology models. In particular, the large deviations caused by the assumed theoretical DMO HMF models will need to be addressed as they could lead to large biases even for current surveys. Although we found that the \citet{Bocquet2016} HMF gives better agreement than popular alternatives, this may be largely coincidental given that we use a different halo finder and that the overestimate of the abundance for masses near the selection limit compensates the strong underestimate for high masses (see Fig.~\ref{fig:HMF_comp}). The divergence between different models for the HMF at high mass suggests that the abundance of these objects might simply be too sensitive to systematic effects, including simulation volume, halo finder and the mass-concentration relation, to give reliable cosmological constraints. Instead it might be favourable to use objects with a lower mass, something that will happen naturally with new surveys. Improvements to the constraints on the observable-mass scaling relation and the effect of baryons on the HMF will also be necessary for next generation surveys.

Two major steps that we have excluded from this work are the measurement of the observable mass proxy and the selection of clusters based on such measurements. Forward modelling using virtual observations would allow including these steps. This would enable application of full multi-frequency SZ selection methods \citep{Melin2006,Melin2012,Hilton2018} to the FLAMINGO lightcones, including foregrounds and noise specific to each survey. Systematic errors in the calibration of the observable-mass relation using e.g.\ gravitational lensing could also be modelled using virtual observations. 

Another avenue worth exploring, is to abandon the classical HMF based on the spherical overdensity definition of dark matter haloes. Instead, we can predict the abundance of clusters as a function of the observable \citep[e.g.][]{Debackere2022a}. This approach can reduce baryonic uncertainties \citep[e.g.][]{Debackere2022b} and is well-suited to emulation. Indeed, given that emulators have already replaced (semi-)analytic models for the DMO HMF, this would be a natural next step. Emulators based on hydrodynamic simulations can directly predict cluster counts as a function of observables such as the limiting Compton-Y parameter and the cosmological parameters. Moreover, they can be used to marginalize over the uncertainties in the baryon physics and to constrain such variations using external data.

\section*{Acknowledgements}
We are grateful to all members of the FLAMINGO collaboration for their help in creating the simulation suite. We thank Ian McCarthy for helpful discussions and the anonymous referee for a constructive report that significantly improved the clarity of the manuscript. For this work we made use of the python packages \texttt{colussus} \citep{Colossuspython}, \texttt{hmf} \citep{HMFpython}, \texttt{corner} \citep{cornerpython} and \texttt{emcee} \citep{emcee2013}. This work is partly funded by research programme Athena 184.034.002 from the Dutch Research Council (NWO). VJFM acknowledges support by NWO through the Dark Universe Science Collaboration (OCENW.XL21.XL21.025). This work used the DiRAC@Durham facility managed by the Institute for Computational Cosmology on behalf of the STFC DiRAC HPC Facility (www.dirac.ac.uk). The equipment was funded by BEIS capital funding via STFC capital grants ST/K00042X/1, ST/P002293/1, ST/R002371/1 and ST/S002502/1, Durham University and STFC operations grant ST/R000832/1. DiRAC is part of the National e-Infrastructure.

\section*{Data Availability}

The code and data used for this work can be found on GitHub\footnote{\href{https://github.com/Moyoxkit/cluster-counts}{https://github.com/Moyoxkit/cluster-counts}}. The HMF and $M_{\rm 500c}$-Compton-Y scaling relation data from FLAMINGO will be shared upon reasonable request to the corresponding author.



\bibliographystyle{mnras}
\bibliography{example} 





\bsp	
\label{lastpage}
\end{document}